\documentclass[a4paper]{article}
\usepackage[margin=25mm]{geometry}

\usepackage[version=4]{mhchem}
\usepackage{bm}
\usepackage{lmodern} 
\usepackage{xifthen}
\usepackage{graphicx}
\usepackage{caption}
\usepackage{subcaption}
\usepackage[T1]{fontenc}
\usepackage{float}
\usepackage{amsfonts}
\usepackage{graphicx}
\usepackage{verbatim}
\usepackage{algorithm}
\usepackage{algorithm,algpseudocode}

\algnewcommand{\LineComment}[1]{\State $\#$ #1}
\usepackage[backend=biber,style=numeric,maxnames=5,minnames=1,maxcitenames=2,uniquename=false,sorting=none]{biblatex}
\newcommand{\citet}[1]{\textcite{#1}}

\newcommand{\pos}[2]{\boldsymbol{#1}_{#2}}
\newcommand{\posbar}[2]{\bar{\boldsymbol{#1}}_{#2}}
\newcommand{\parentheses}[3]{\left#1#2\right#3}

\newcommand{\prob}[3][]{\ifthenelse{\isempty{#1}}{P\left(#2,#3\right)}{P\left(#1,#2,#3\right)}}
\newcommand{\probGiven}[4][]{\ifthenelse{\isempty{#1}}{P\parentheses{(}{#2,#3|#4}{)}}{P\parentheses{(}{#1,#2,#3|#4}{)}}}

\newcommand{\probbar}[3][]{\ifthenelse{\isempty{#1}}{\bar{P}\left(#2,#3\right)}{\bar{P}\left(#1,#2,#3\right)}}

\title{Particle-based simulation of non-elementary bimolecular kinetics}
\author{Taylor Kearney and Mark B. Flegg}
\date{\today}

\begin{document}

\maketitle

\section*{Abstract}

Particle-based simulations are an essential tool for the study of biochemical systems for scales between molecular/Brownian dynamics and the reaction-diffusion master equation. These simulations utilise proximity-based reaction conditions and are typically limited to elementary (mass-action) kinetics. We present a novel framework for directly simulating non-elementary bimolecular kinetics in a particle-based framework. By mimicking the behaviour of a third implicit reactant, we adapt non-elementary reaction conditions, previously restricted to trimolecular chemical interactions, to biomolecular reactions for the first time. We implement our approach in an event-driven simulation, which we validate by reproducing Michaelis-Menten kinetics. We then demonstrate its utility by simulating the classical Goldbeter model of circadian oscillations completely at the level of individual molecules. This model features multiple non-elementary reactions and requires the incorporation of several existing simulation techniques. Our method accurately reproduces the target non-elementary kinetics, without simulating the implied underlying fast elementary reactions, thereby significantly reducing the computational cost. This work expands the class of reaction networks accessible to particle-based simulations and provides a practical alternative to explicitly simulating all elementary steps in systems where quasi-steady-state approximations are applicable.

\section{Introduction}

The behaviour of biological systems is governed by the integrated function of numerous processes that operate across a variety of distinct spatiotemporal scales. At molecular scales, individual molecules span only a few nanometres and can evolve in just picoseconds \cite{ZWIER2010745}. Ultimately, these subcellular molecular interactions govern the behaviour of entire multicellular organisms that might live for years and measure a metre or more. This vast range of scales, combined with the inherent complexity of biological systems, makes modelling them exceptionally challenging, and detailed simulations of large organisms remain beyond the capabilities of current hardware.\par 

Even the more modest goal of developing a computational model of a single cell has been labelled a `grand challenge' for systems biology \cite{TOMITA2001205, harel2003grand, harel2005turing}. Nonetheless, considerable progress has been made in the development of whole-cell models, which aim to simulate the integrated function of every gene within a cell \cite{TOMITA2001205, GOLDBERG201897}. For example, \citet{karr2012whole} constructed a whole-cell model of the simple organism Mycoplasma genitalium \cite{mycoplasma}, and more recently, whole-cell models have been developed for more complex singled-celled organisms such as E. coli \cite{ecoli_og, ecoli1, ecoli2} and Saccharomyces cerevisiae (Brewer's yeast) \cite{Saccharomyces_cerevisiae}.\par

Biology is fractal by nature, and the challenges involved in modelling multicellular organisms are shared by whole-cell models. Essential cellular functions, such as metabolism and signalling, are governed by long, interconnected reaction pathways that involve numerous chemical species---mainly RNA and protein molecules---and a multitude of reactions \cite{BHALLA200345, VISWAN2023102808}. The reaction rates within these pathways can vary wildly. A single protein molecule undergoes approximately one diffusion-driven collision per mole of other reactants every nanosecond, and the fastest catalysed intracellular reactions occur in just microseconds. In contrast, the slowest catalysed reactions can take up to a minute, while the degradation of mRNA and proteins within human cells can span hours to days \cite{bio_numbers}.\par 

To combat the computational challenges posed by this range of scales, whole-cell models favour modular designs that enable them to incorporate multiple approaches \cite{SOUTHERN200860, 10.1039/c0ib00075b}, each operating with varying levels of spatiotemporal resolution and targeting specific cellular processes \cite{multiscale_review}. In principle, this allows whole-cell models to prioritise their limited computational resources by utilising accurate, and typically costly, simulation methodologies only where necessary. However, in practice, the most accurate methodologies are rarely used owing to their computational intensity. Instead, whole-cell models are forced to rely on computationally efficient methods such as flux balance analysis (FBA) \cite{OrthJeffreyD2010Wifb}---which is often used to model cellular metabolism \cite{karr2012whole, ecoli_og}---and deterministic rate equations (REs), which describe the evolution of time-dependent, mean-field concentrations \cite{chen2010classic}. However, the reliance on these models fundamentally limits the accuracy of whole-cell models.

The prevalence of REs in particular, in whole-cell models and biochemical modelling more widely, is predicated on the assumption that reactants in cellular systems can be approximated using well-mixed deterministic concentrations. The assumption of well-mixedness requires that reactant diffusion is fast enough that the distance that any particular biomolecule is expected to travel during its lifetime (the time before it is consumed by a reaction), called the Kuramoto length \cite{Kuramoto_length}, is significantly larger than the length of the reactive volume \cite{PhysRevE.70.020901}. Moreover, the assumption of determinism requires that stochastic fluctuations, whose importance in cellular systems has been widely reported \cite{raj2008nature, E_coli_final, 10.1371/journal.pcbi.1002010,doi:10.1126/science.1147888, stochastic_processes_bio_book}, can be neglected. The magnitude of these fluctuations for a given reactant scales with the square root of the number of molecules \cite{Kuramoto_length, chen2010classic}. Therefore, determinism requires large molecular populations. This assumption appears to be valid at least superficially, for many systems and a single E. coli cell for example, contains approximately two million proteins \cite{protein_count} with an average protein copy number of approximately $3.5\times 10^3$ \cite{e_coli_abundance}. However, the mean is skewed by the fact that some proteins are very abundant and \citet{e_coli_abundance} report a median protein copy number in E. coli of 526, with $75\%$ and $25\%$ of the proteins they identified occurring with counts less than 1160 and 250, respectively. Therefore, it is not possible to define a satisfactory concentration for many vital biochemical components, either because they occur in very low numbers or because they are spatially localised \cite{erban2009stochastic}.\par 

Fundamentally, stochasticity in cellular kinetics arises as a result of the inherently random motions of individual molecules \cite{SHAHREZAEI2008369}. Thus, a natural approach, adopted by molecular dynamics (MD) simulations \cite{molecular_dynamics_simulations}, is to explicitly model the behaviour of reactants at an atomic level. Unfortunately, achieving atomic precision is incredibly computationally intensive, and despite the proliferation of powerful commercially available graphics processing units \cite{GPU_MD_review}, and the rapid progress achieved by leveraging high-performance computing \cite{10.1145/3503221.3508425,doi:10.1177/10943420221128233}, it is still not possible to simulate biologically relevant timescales with this level of precision \cite{hollingsworth2018molecular}. \citet{10.3389/fchem.2023.1106495} predict that a coarse-grained MD simulation, where a single particle represents a group of atoms \cite{coarse_grained_MD}, of the minimal Mycoplasma bacterium called \textit{JCVI-syn3A} \cite{doi:10.1126/science.aad6253} may be feasible in the near future. However, even this minimal model contains $5.6 \times 10^8$ individual elements and although the predicted simulation of 10–100$\mu$s of  evolution would be incredibly impressive, it is still orders of magnitude smaller than relevant timescales for all but the fastest biological processes.\par

Current whole-cell models \cite{modern_whole_cell_review, ecoli2} typically incorporate stochasticity, if at all, using the stochastic simulation algorithm (SSA) \cite{GILLESPIE1976403, Gillespie1977} which is a computationally efficient method for sampling realisations of the chemical master equation. However, the CME still assumes that the reactants are well mixed on the scale of the domain and is devoid of any spatial information. To rectify this, the domain can be divided into small compartments or voxels so that the reactants only need to be considered well mixed within individual voxels \cite{PhysRevE.71.041103, Gillespie_spatial_inhomo}. This leads to the reaction-diffusion master equation (RDME) \cite{Isaacson_RDME}, where the CME is applied locally to each voxel and reactant motion is modelled using diffusive transfers between neighbouring voxels.\par

The spatial resolution of the RDME is limited by the size of its voxels and it becomes impractical when reactants occur with low copy numbers because many compartments are required to localise a small number of molecules. Moreover, because reactions are only allowed to occur between reactants within the same voxel, it does not converge in the limit that the voxels become infinitesimal because bimolecular reactions vanish \cite{PhysRevE.85.042901}. Under such circumstances, it is preferable to use particle-based models that explicitly track individual molecules of the relevant reactants in the vein of MD simulations. Crucially, though, particle-based models make several simplifying assumptions, examined in detail by \citet{Smith2019}, that make them orders of magnitude more efficient than MD simulations. In particular, they do not explicitly model solvent molecules, and instead incorporate them implicitly by supposing that reactant molecules undergo independent isotropic diffusion as a result of their collisions with these implicit molecules. In addition, they typically do not account for external or intermolecular forces, and reactions are modelled using simple proximity-based reaction conditions that amount to supposing that a reaction occurs when two reactants `collide' at a predefined distance known as the \textit{reaction radius}. This approach was largely inspired by Smoluchowski \cite{smoluchowski1917versuch} and forms the basis of many prominent software packages for particle-based simulation of reaction-diffusion systems, including: ChemCell \cite{chemcell}, MCell \cite{MCell1, MCell2} Green’s Function Reaction Dynamics (GFRD) \cite{gfrd1,gfrd2}, enhanced Green's Function Reaction Dynamics (eGFRD) \cite{egfrd1,egfrd2}, and Smoldyn \cite{Andrews_2004}.\par

Particle-based simulations are widely used in the literature \cite{schoneberg2014simulation} and can be used to simulate the behaviour of reaction networks over macroscopic timescales using personal computing hardware. They offer a high degree of spatial resolution and explicitly reproduce stochasticity as the result of random intermolecular collisions. Nonetheless, they are not immune to the computational challenges posed by intracellular complexity. In particular, explicitly tracking all the molecules within a cell is infeasible on current hardware and they are seldom incorporated even to simulate small subsystems in whole-cell models.\par

Traditionally, particle-based simulations of biochemical pathways are constructed by explicitly modelling each of the constituent elementary reactions. Unimolecular reactions are assumed to occur instantaneously and can be modelled as Poissonian processes that are independent of molecular diffusion \cite{Andrews_2004,egfrd2}, while bimolecular reactions are governed by proximity-based reaction conditions---akin to Smoluchowski's, or one of its many derivatives \cite{collins1949diffusion, Doi_1976, erban2009stochastic}---and molecular diffusion. Whenever there is potential for a reaction to occur, the exact positions of the relevant molecules must be known so that the reaction condition can be tested. Therefore, the majority of the computational effort in particle-based simulations is dedicated to updating molecular positions and then determining molecular separations. The frequency of such updates is dictated by the reaction rate and resolving fast reactions requires very short timescales.\par

Biochemical reactions are often catalysed by a class of proteins known as \textit{enzymes}, which selectively accelerate specific reactions. The products produced at each stage typically form the input for the next reaction, resulting in intricate pathways with multiple interconnected steps \cite{cell_book}. Due to enzyme activity, these pathways operate across disparate timescales, making their particle-based simulation computationally expensive. However, these timescales can be exploited to construct reduced models \cite{snowden2017methods} that focus on essential subsystems or dominant interactions \cite{NFKB, MENGEL2010656, MAPK,10.1371/journal.pcbi.1001004}, omitting intermediates that may influence kinetics, but are not of particular interest to the modeller. These simplifications allow for clearer interpretation and reduce the number of rate parameters required, which is beneficial because our knowledge of these parameters—--particularly in vivo—--remains incomplete \cite{doi:10.1021/bi2002289, heckmann2018machine}.\par 

By exploiting timescale separation, it is possible to eliminate fast or intermediate steps, thereby simplifying the system while preserving the essential dynamics. In principle, this approach can be used to improve the efficiency of particle-based simulations. However, such reductions often result in non-elementary reactions that proceed at rates that cannot be described by the law of mass action and as a result cannot be implemented using standard proximity-based reaction conditions.\par 

The prototypical example of such a reduction was first performed by \citet{michaelis1913kinetik} who, building upon the earlier work of \citet{henri1903lois}, considered an enzymatic system consisting of a substrate ($S$) that binds to an enzyme ($E$) to produce an intermediate bound chemical complex ($C$). Once in the bound state $C$, the substrate can be converted to a product, freeing the enzyme to bind with another substrate molecule, or the complex can disassociate into its constituent components. Because we are only interested in the rate at which this reaction proceeds, we can ignore the product and instead only consider the degradation of the substrate. That is, we consider the reaction
\begin{equation}\label{eq:MM_network}
    \ce{S + E <=>[k_1][k_{-1}] C ->[k_2] E},
\end{equation}
where the second-order rate constant $k_1$ controls the rate of formation of $C$, while $k_{-1}$ and $k_2$ are the first-order rate constants that control the rate of dissociation of $C$ and the degradation of bound substrate, respectively \cite{CORNISHBOWDEN19791}.\par

The traditional analysis, more rigorously justified by \citet{A_note_on_the_kinetics_of_enzyme_action} than Michaelis and Menten, of this system adopts the quasi-steady-state approximation (QSSA) which amounts to assuming that on the timescale over which significant substrate degradation occurs, the substrate molecules bind to free enzymes so rapidly that the amount of $C$ is always in an instantaneous steady state with the amount of $S$; a so called quasi-steady-state \cite{the_quasi-steady-state_assumption}. The simplifications enabled by this assumption then yield the well-known Michaelis-Menten equation which describes the degradation rate of $S$,
\begin{equation}\label{eq:MM_production_rate}
    \frac{ds}{dt} = -\frac{k_2e_0s}{K_m+s},
\end{equation}
where $K_m = (k_{-1} + k_2)/k_1$ is the Michaelis constant, $e_0$ denotes the initial concentration of the unbound enzyme and $s$ denotes the concentration of the substrate at time $t$. Although often stated verbatim, the QSSA arises from applying singular perturbation theory to the system of ordinary differential equations (ODEs) that describes Reaction (\ref{eq:MM_network}) \cite{HEINEKEN196795} and only requires that the system contains sufficiently disparate timescales \cite{the_quasi-steady-state_assumption} making it applicable to much more complicated systems than the one considered by Michaelis and Menten.\par

Now consider constructing a particle-based simulation for Reaction (\ref{eq:MM_network}). The usual approach requires explicitly simulating the formation of intermediary $C$, which means we must simulate the very rapid binding events between the substrate and enzyme. Thus, reproducing the macroscopic behaviour of the system becomes computationally intensive since the simulation must operate on a timescale that is short enough to resolve these binding events, which is necessarily much shorter than timescale over which significant degradation of the the substrate occurs. However, if we are primarily concerned with the degradation of the substrate (or its conversion into some product) then the preceding analysis offers a way to circumvent this issue. Namely it describes the long term behaviour of the system, without reference to the complex $C$, using a single reaction between the substrate and enzyme which does not require us to resolve individual enzyme and substrate bindings. In essence, it enables us to incorporate the action of the intermediary $C$ without requiring this complex, or its formation, to be explicitly modelled. Therefore, this analysis not only reduces the number of chemical species we must track, but more importantly, drastically reduces the number of reactions that must be simulated and significantly increases the size of the timescale that needs to be resolved to attain the desired kinetics.\par

To take advantage of this analysis we must be able to reproduce the non-elementary kinetics described by Equation (\ref{eq:MM_production_rate}) in a particle-based simulation, without explicitly simulating Reaction (\ref{eq:MM_network}). Instead we model just a single reaction
\begin{equation}\label{eq:direct_MM_reaction}
    \ce{S + E ->[k(s)] E}
\end{equation}
between the substrate and enzyme that directly describes the degradation of the substrate, where the rate parameter $k(s) = k_2/(K_\mathrm{M} + s)$ depends explicitly on the substrate concentration $s$. Unfortunately, applying Smoluchowski's reaction condition, or one of its various derivatives \cite{collins1949diffusion, Doi_1976}, to Reaction (\ref{eq:direct_MM_reaction}) only ever yields mass-action kinetics, where the corresponding rate parameter is always a constant that depends on the properties of the reaction condition, but never the concentration $s$. Therefore, we require a bimolecular proximity-based reaction condition that reproduces the non-elementary kinetics of Reaction (\ref{eq:direct_MM_reaction}).\par

In their recent work, \citet{our_first_paper} demonstrate that non-elementary reaction rates can be reproduced in particle-based simulations using a proximity-based reaction condition. However, their condition only applies to trimolecular systems and relies on an extra spatial degree of freedom that does not exist in bimolecular systems. Specifically, Kearney and Flegg's condition depends not only on the pairwise proximity between two molecules but also on their relative proximity to the closest molecule of the third species. In bimolecular systems, only the first of these proximities can be defined, preventing direct application of their method.\par 

In this article, we address this limitation by developing a particle-based framework that extends Kearney and Flegg's reaction condition to bimolecular systems. This allows us to reproduce non-elementary kinetics in bimolecular systems, without explicitly simulating the underlying fast elementary reactions, thereby improving the efficiency of particle-based simulations of enzymatic systems. Notably, this behaviour cannot be replicated by the RDME, the closest alternative to particle-based models \cite{Smith2019}, because it is not guaranteed to converge when non-elementary propensities are used \cite{RDME_MM_nonconvergence}. In addition, the accuracy of the Michaelis-Menten equation in stochastic settings is known to reduce with decreasing system volume \cite{RDME_MM_volume_error}. While this is often negligible at biologically relevant scales, the subdivision of the volume required by the RDME introduces significant error at scale of individual voxels. This error can be reduced by increasing the size of the voxels, but doing so sacrifices spatial resolution.  \par 

We begin in Section~\ref{sec:Review} by summarising the results of \citet{our_first_paper} that underpin our framework, before showing how to adapt these results to bimolecular systems in Section~\ref{sec:modelling_bimolecular_systems}. Section~\ref{sec:simulation_details} outlines the implementation of our method within an event-driven particle-based simulation, which we then use to generate the results presented in Section~\ref{sec:results}. In particular, in Section~\ref{subsec:MM_results} we validate our approach by reproducing Michaelis-Menten kinetics, while in Section~\ref{subsec:oscillations} we demonstrate how our method can be incorporated with existing techniques by simulating a model of circadian rhythms developed by \citet{og_circadian_model}.\par

\section{Reaction conditions for non-elementary kinetics}\label{sec:Review}
Our approach builds upon the mathematical framework developed by \citet{flegg2016smoluchowski} and later extended by \citet{our_first_paper}. This framework generalises proximity-based reaction conditions to reactions between an arbitrary number of reactants and provides a convenient way to quantify the relative proximity between multiple diffusing points. Kearney and Flegg previously used this framework to demonstrate that non-elementary kinetics, resembling Michaelis-Menten kinetics, can be modelled using proximity-based reaction conditions in trimolecular systems. Although we focus solely on bimolecular reactions and consider only two distinct chemical species in our particle-based simulations, it is instructive to begin with a trimolecular system akin to that of Kearney and Flegg. Thus, we will first review how to reproduce kinetics analogous to those of Reaction (\ref{eq:MM_network}) using a trimolecular reaction condition, and then introduce a modification in Section~\ref{sec:modelling_bimolecular_systems} that allows analogous conditions to be applied to bimolecular systems.\par

We consider a system in which a single molecule of the enzyme $E$ is surrounded by $N_S = sV$ molecules of the substrate $S$, where $s$ is a well-mixed concentration of substrate molecules and $V$ is the volume of the domain, which is finite but very large. In addition, the system contains a single molecule of a third chemical species labelled $X$. We assume that all the molecules diffuse independently within the domain. The system contains $N_S$ distinct states, where each state contains the enzyme molecule, the molecule of $X$, and a particular substrate molecule.\par

To apply a proximity-based reaction condition to this system, we must be able to quantify the relative proximity of the constituent molecules for any of the $N_S$ distinct states. To understand how this is achieved, we momentarily restrict our attention to just one of these states. For this particular state, we will use $D_i$ and $\mathbf{x}_i$ where $i = 0, 1, 2$, to denote the diffusion constant and the $3$-dimensional position of the molecules of $E$, $X$ and $S$, respectively. We then adopt the diffusive Jacobi coordinates or separation coordinates \cite{flegg2016smoluchowski},
\begin{subequations}
\begin{align}
    \pos{\eta}{0} &= \posbar{x}{2} \quad \text{and} \quad \label{eq:eta1_def}\\
    \pos{\eta}{i} &= \pos{x}{i} - \posbar{x}{i-1}, \quad i = 1,2,  \quad \text{where,}\label{eq:etai_def}\\
    \posbar{x}{i} &= \frac{\sum^i_{j=0} \pos{x}{j}D_j^{-1}}{\sum^i_{k=0}D_k^{-1}},\label{eq:x_bari_def}
\end{align}\end{subequations}
is the \textit{centre of diffusion} of the first $i$ molecules, which is analogous to the centre of mass except that the positions are weighted by their inverse diffusion coefficients rather than their masses.\par 

\begin{figure}
\centering
\includegraphics[width=0.5\textwidth]{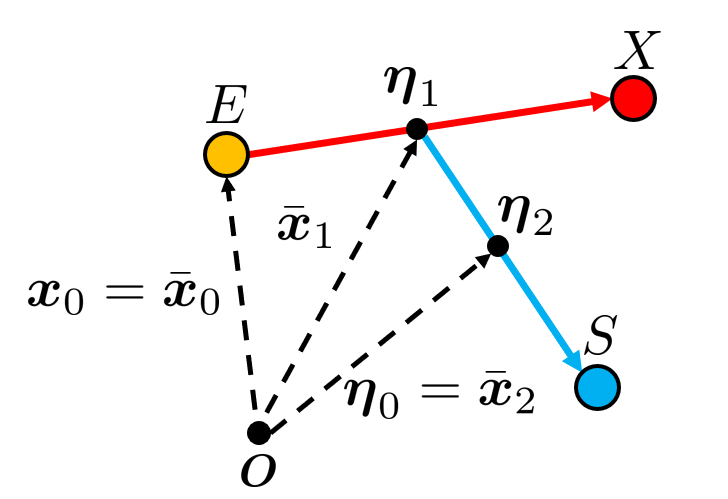}
\caption{The diffusive Jacobi coordinates that quantify the relative proximity for a triplet of molecules. The coordinate $\pos{\eta}{1}$ describes the separation between the first two molecules, labelled $E$ and $X$, while $\pos{\eta}{2}$ describes the separation of the third molecule $S$ from the centre of diffusion, denoted $\posbar{x}{1}$ and defined in Equation (\ref{eq:x_bari_def}), of $E$ and $X$. Finally, $\pos{\eta}{0}$ is the centre of diffusion of the three molecules ($\posbar{x}{2}$) and changes in this coordinate correspond to translations of the entire triplet in space.}
\label{fig:eta_coordinates}
\end{figure}

In this coordinate system, the molecules' relative proximity is captured by $\pos{\eta}{1}$ and $\pos{\eta}{2}$, which describe, respectively, the separation between $E$ and $X$, and the separation between $S$ and the centre of diffusion, $\posbar{x}{2}$, of $E$ and $X$; see Figure \ref{fig:eta_coordinates}. Meanwhile, $\pos{\eta}{0}$ corresponds to the centre of diffusion of the three molecules and translations in this coordinate do not change their relative proximity. Therefore, reaction conditions are independent of $\pos{\eta}{0}$ and depend only on $\pos{\eta}{1}$ and $\pos{\eta}{2}$. In addition, $\pos{\eta}{1}$ and $\pos{\eta}{2}$ undergo independent linear diffusion with the respective diffusion constants
\begin{subequations}
  \begin{align}\label{eq:eta_i_diff_coeff}
    \hat{D}_1 &= D_{1} + \bar{D}_{0},\\
    \hat{D}_2 &= D_{2} + \bar{D}_{1},  \quad \text{where}\\
    \bar{D}_j &= \frac{1}{\sum_{i=0}^{j} D_i^{-1}},
   \end{align}  
\end{subequations}
is the diffusion constant associated with $\posbar{x}{j}$.\par

We can now define a general proximity-based reaction condition for the system as the absorbing boundary $\partial \Omega$
\begin{equation}\label{eq:reactive_domain}
    \partial \Omega = \left\{(\pos{\eta}{1}, \pos{\eta}{2}) : r_1 = f(r_2)\right\},
\end{equation}
where $r_i = ||\pos{\eta}{i}||$ is the radial coordinate associated with the position $\pos{\eta}{i}$, $0 \leq f(r_2) \leq \sigma$ is a non-negative function of $r_2$ and $\sigma$ is a small positive constant relative to the diffusion coefficients $\hat{D}_i$ and the characteristic scale of the domain of $f$ \cite{our_first_paper}. The system only contains a single pair of $E$ and $X$ molecules, so the distance $r_1$ between them is unambiguous. However, because the system contains multiple molecules of $S$, the value of $r_2$ depends on the particular molecule of $S$ under consideration. That is, the reaction condition and the resulting kinetics depend on the particular state used to test the condition.\par 

If we restrict our attention to the molecule of $S$ that is closest to the centre of diffusion of $E$ and $X$, then the steady-state reaction rate $K(s)$ is given by the total flux of the steady-state probability density $P(\pos{\eta}{})$ to find the state with the minimum value of $r_2$ at position $\pos{\eta}{} = \{\pos{\eta}{1}, \pos{\eta}{2}\}$. Determining this flux analytically is very difficult, but \citet{corrections_paper} show that to leading order in $\sigma$ it is given by the diffusive flux of $P$ over $\partial \Omega$ in the $\pos{\eta}{1}$ direction
\begin{equation}\label{eq:closest_rate_old}
    K_1(s) = \int_0^\infty \frac{4\pi \hat{D}_1 f(r_2)}{V}s\text{exp}\left(\frac{-4\pi sr_2^3}{3}\right)4\pi r_2^2 \,\mathrm{d}r_2.
\end{equation}
The label $K_1(s)$ is used to distinguish this approximate rate from the exact rate $K(s)$ associated with $\partial \Omega$ since it neglects corrections of $O(\sigma^2)$ to the flux over $\partial \Omega$. Specifically, \citet{corrections_paper} show that $K_1(s)$ neglects the diffusive and advective fluxes of $P$ over the boundary in the $\pos{\eta}{2}$ direction, as well as a correction to the diffusive flux in the $\pos{\eta}{1}$ direction. The neglected advective flux arises because there is an advection towards $\pos{\eta}{2}=\pos{0}{}$ due to the fact that the molecule of $S$ with the second-smallest value of $||\pos{\eta}{2}||$ can diffuse past the currently closest molecule of $S$ to become the new closest molecule that defines $r_2$.\par 

Nonetheless, Equation (\ref{eq:closest_rate_old}) is useful because it provides an analytic expression that can be solved, via an inverse Laplace transform, for the function $f$ that defines the reaction boundary for a given reaction rate $K_1(s)$. Moreover, once the analytic form of $\partial \Omega$ has been determined, it can be altered so as to minimise the error between the reaction rate it reproduces and the original desired rate \cite{corrections_paper}. In other words, because Equation (\ref{eq:closest_rate_old}) is approximate, using it to derive $\partial \Omega$ will result in a reaction rate that differs from what is expected, but this boundary can be corrected to minimise this error.\par

\citet{our_first_paper} first demonstrated that non-elementary kinetics can be reproduced directly in a particle-based simulation using the reaction condition
\begin{equation}\label{eq:MM_reaction_condition}
    \partial\Omega_{\mathrm{MM}} = \left\{(\pos{\eta}{1},\pos{\eta}{2}) : r_1 = \sigma \mathrm{Exp}\left(\frac{-4\pi\Gamma r_2^3}{3}\right) \right\},
\end{equation}
which yields a reaction rate $K_1(s) = 4\pi\hat{D}_1\sigma s/(V(\Gamma + s))$ that resembles the reaction rate of Reaction (\ref{eq:direct_MM_reaction}). Note that the reaction rate $K(s)$ is the number of reactions per unit time for a single pair of $E$ and $X$ molecules. However, this trimolecular condition requires two spatial degrees of freedom, $r_1$ and $r_2$. In contrast, a bimolecular system offers only a single spatial degree of freedom, the distance between two molecules within each pair, with which to define the reaction condition. Therefore, it is not clear how to apply this reaction condition to bimolecular reactions, preventing us from directly simulating Reaction (\ref{eq:direct_MM_reaction}). Fortunately, it is possible to introduce a simple mechanism that reintroduces a second spatial degree of freedom, and allows us to use reaction conditions of the form given in Equation (\ref{eq:reactive_domain}) in bimolecular systems.


\section{Modelling bimolecular systems}
\label{sec:modelling_bimolecular_systems}
Adapting the reaction conditions proposed in Section~\ref{sec:Review} to bimolecular systems requires replicating the behaviour of the additional reactant available when designing trimolecular conditions. That is, we aim to mimic the behaviour of the aforementioned trimolecular system after removing the reactant $X$ so that the system contains just a single enzyme molecule surrounded by substrate molecules. In this bimolecular system it is still possible to identify the closest molecule of $S$ to $E$ at any given moment and hence recover a distance analogous to $r_2$. However, there is no obvious way to define $r_1$, preventing application of reaction conditions of the form given in Equation (\ref{eq:reactive_domain}). Before discussing how to reintroduce this spatial degree of freedom in a manner consistent with the trimolecular system, it is helpful to first quantify the behaviour we are trying to replicate.\par 

\begin{figure}
\centering
\includegraphics[width=0.4\textwidth]{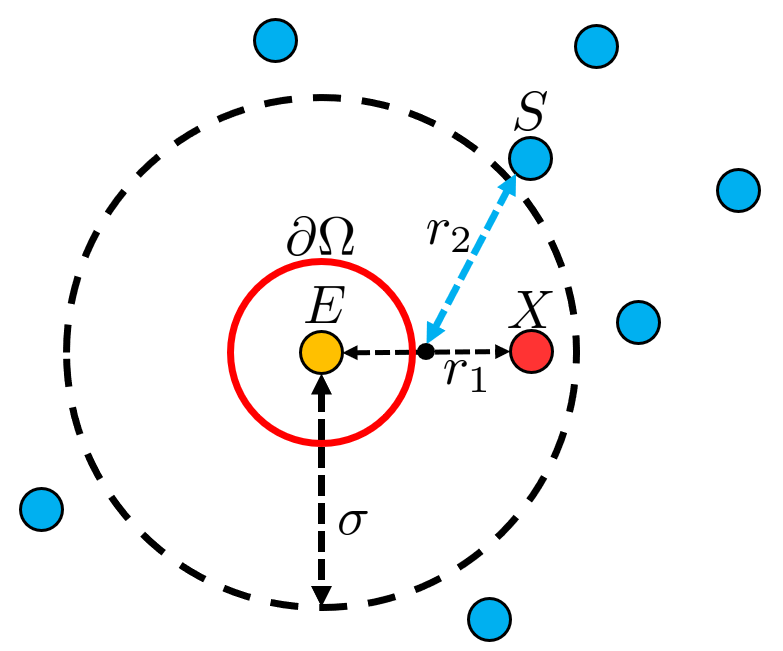}
\caption{The absorbing reaction boundary $\partial \Omega$ in Equation (\ref{eq:reactive_domain}) can be viewed as defining a spherical surface, shown by the red circle, that is centred on the molecule of $E$, represented by the yellow point. This boundary attains a maximum radius of $\sigma$, and any time the radial distance $r_1$ between the molecule of $E$ and the molecule of $X$ (the red point) is such that $r_1 \leq \sigma$, the pair is considered reactive. However, a reaction only occurs when the molecule of $X$ reaches $\partial \Omega$, whose radius $f(r_2)$ is a function of the distance $r_2$ from the centre of diffusion of the molecules of $E$ and $X$ to the closest molecule of $S$, depicted by the blue points.}
\label{fig:reaction_boundary}
\end{figure}

Equation (\ref{eq:reactive_domain}) dictates that a reaction occurs when the distance $r_1$ between the molecules of $E$ and $X$ is equal to $f(r_2)$, where $r_2$ is the distance from the centre of diffusion of $E$ and $X$ to the closest molecule of $S$ within the domain. The parameter $\sigma$ determines the maximum value of $r_1$ for which a reaction can occur, so that if $r_1 > \sigma$ no reaction is possible and the reaction condition does not need to be checked. However, whenever $r_1 \leq \sigma$ then a reaction is possible, and we say that the pair of $E$ and $X$ molecules is in a \textit{reactive} state. That is, we consider the pair to be reactive whenever their separation is such that it is possible for a reaction to occur. Whether the reaction actually occurs is determined by the relative proximity of the closest molecule of $S$ and requires the relative position of the three molecules to be such that $r_1 = f(r_2)$ at any point before $E$ and $X$ diffuse away from each other. In this way, the distance to the closest molecule of $S$ can be viewed as mediating the affinity the molecules of $E$ and $X$ have for each other, as shown in Figure~\ref{fig:reaction_boundary}.\par 

The rate at which the molecules of $E$ and $X$ become reactive---that is, the rate at which these molecules approach at a distance $\sigma$---is given by the Smoluchowski bimolecular reaction rate \cite{smoluchowski1917versuch} associated with the reaction radius $\sigma$
\begin{equation}\label{eq:smol_bimolecular_rate}
    K_\text{R} = \frac{4\pi \hat{D}_1\sigma}{V}.
\end{equation}
As a consequence, the times at which the pair of $E$ and $X$ molecules become reactive are exponentially distributed with mean $K_R$ such that
\begin{equation}
\label{eq:AB_arrival_times}
    P_\mathrm{R}(t) = K_\text{R} \text{ exp}\left(-K_\text{R}  t\right),
\end{equation}
where $P_\mathrm{R}(t)$ is the probability density for the waiting time for entering into a reactive state between $E$ and $X$ (whereby $E$ and $X$ are initially separated by $r_1 \leq \sigma$). We do not typically concern ourselves with the position of $X$. Instead, what we take from this is only that the arrival times of $X$ on a sphere of radius $\sigma$ around $E$ are distributed according to Equation (\ref{eq:AB_arrival_times}).\par 

When the molecule of $E$ becomes reactive, we check the likelihood that the full reaction condition is met. Specifically, we determine whether the reactive pair dissociates ($r_1 \rightarrow \infty$) or a reaction occurs ($r_1 = f(r_2)$) first. Since $\sigma$ is small, $X$ explores the reactive volume about $E$ rapidly, and the positions of the $S$ molecules can be considered to be effectively static during this period. Therefore, to determine whether a reaction should occur, it is sufficient to compare the value of $f(r_2)$ at the time the pair becomes reactive to the minimum value of $r_1$ attained before the pair dissociates (the point of closest approach).\par

\begin{figure}
\centering
\includegraphics[width=0.4\textwidth]{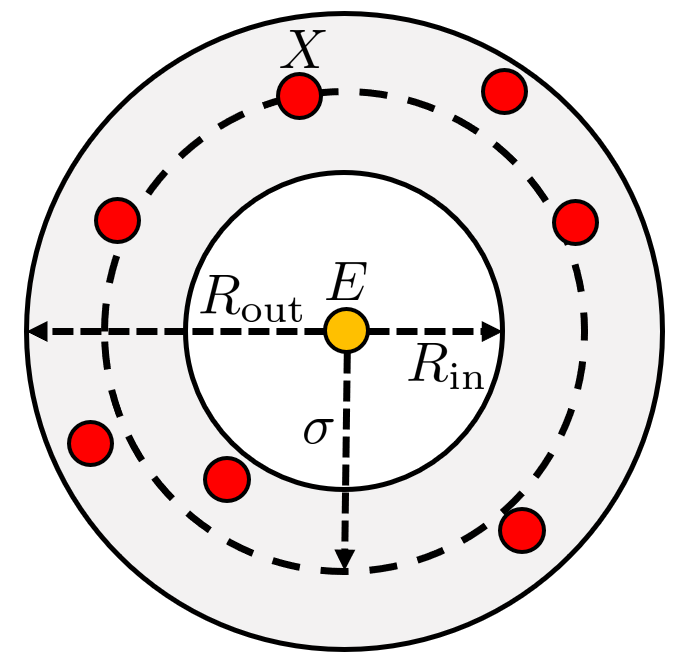}
\caption{To determine the smallest radial distance between two diffusing molecules $E$, the yellow point, and a molecule of $X$ that are initially separated by the distance $\sigma$, we imagine that $E$ is the centre of three concentric spheres of radii $R_{\text{in}}$,  $\sigma$, and $R_\text{out}$, where $R_{\text{in}} < \sigma < R_\text{out}$. Molecules of $X$, shown by the orange points, are produced on the surface of the sphere of radius $\sigma$ and are absorbed on the surface of the spheres of radius $R_{\text{in}}$ and $R_{\text{out}}$. We consider the continual production of $X$ molecules since we require the the steady-state probability density for finding a molecule of $X$ with a given radial separation to be non-zero.}
\label{fig:closest_approach}
\end{figure}

To determine the point of closest approach between the molecules, suppose that the molecule of $E$ is the centre of three concentric spheres of radii $R_{\text{in}}$, $\sigma$ and $R_{\text{out}}$, respectively, such that $R_{\text{in}} < \sigma < R_{\text{out}}$. Moreover, assume that on the surface of the sphere of radius $\sigma$, molecules of $X$ are continually produced, which diffuse independently and are then absorbed on the surface of the other two spheres as shown in Figure~\ref{fig:closest_approach}. The continual production of $X$ molecules is assumed so that the steady-state probability density $P_X\left(r_1\right)$ for finding a molecule of $X$ with a radial separation $r_1$ from $E$, is non-zero and evolves according to the diffusion equation
\begin{subequations}
\begin{align}\label{eq:diffusion_closest_approach}
\hat{D}_1\left(\frac{d^2 P_X\left(r_1\right)}{d 
 r_1^2} + \frac{2}{r_1}\frac{d P_X\left(r_1\right)}{d 
 r_1}\right) + \frac{\delta\left(r_1 - \sigma\right)}{4\pi r_1^2} &= 0 \quad \text{with,}\\     
    P_X\left(R_{\text{in}}\right) &= 0 \quad  \text{and} \label{eq:diffusion_closest_approach_IB}\\
    P_X\left(R_{\text{out}}\right) &= 0.\label{eq:diffusion_closest_approach_OB}
\end{align}
\end{subequations}
The solution to Equation (\ref{eq:diffusion_closest_approach}) subject to the absorbing boundary conditions in Equations (\ref{eq:diffusion_closest_approach_IB}) and (\ref{eq:diffusion_closest_approach_OB}) is
\begin{equation}
    P_X(r_1) = \begin{cases}
        \frac{\left(r_1-R_{\text{in}}\right)\left(\sigma-R_{\text{out}}\right)}{4\pi\hat{D}_1r_1\sigma\left(R_{\text{out}}-R_{\text{in}}\right)}, \quad R_{\text{in}} \leq r_1 < \sigma\\~\\
        \frac{\left(r_1-R_{\text{out}}\right)\left(\sigma-R_{\text{in}}\right)}{4\pi\hat{D}_1r_1\sigma\left(R_{\text{out}}-R_{\text{in}}\right)}, \quad \sigma \leq r_1 \leq R_{\text{out}}.
    \end{cases}
\end{equation}\par

The probability that a particular molecule of $X$ reaches a radial distance $r_1 \leq R_{\text{in}}$ before diffusing away, is equal to the probability that the same molecule is absorbed on the surface of the innermost sphere at $r_1 = R_{\text{out}}$, rather than on the surface of the outer sphere at $r_1 = R_{\text{out}}$. This probability is given by the ratio of the diffusive flux of $P_X\left(r_1\right)$ over the inner boundary to the total diffusive flux over both the inner and outer boundaries. Assuming $\sigma$ and $R_{\text{out}}$ are fixed, this ratio is also the probability
\begin{equation}\label{eq:closest_mol_CDF}
    \Phi\left(R_{\text{in}}\right) = \frac{R_{\text{in}}\left(R_{\text{out}}-\sigma\right)}{\sigma\left(R_{\text{out}}-R_{\text{in}}\right)},
\end{equation}
for finding $r_1 \leq R_{\text{in}}$.\par


To determine the minimum value of $r_1$ attained before the reactive pair reaches a separation of $R_{\text{out}}$, we first note that $\Phi\left(r_1\right)$ as defined by Equation (\ref{eq:closest_mol_CDF}) is the cumulative distribution function (CDF) for the distribution of the minimal $r_1$ (the closest approach $r^{\text{min}}_1$). For our purposes we need to track the molecule of $X$ until it has moved sufficiently far from the molecule of $E$. We take the limit as $R_{\text{out}} \to \infty$ in Equation (\ref{eq:closest_mol_CDF}) which is equivalent to $X$ leaving the inner boundary layer near $E$. The CDF for the closest approach $r^{\text{min}}_1$ is therefore
\begin{equation}\label{eq:closest_mol_CDF_infty}
    \Phi_{\infty}\left(r^{\text{min}}_1\right) = \lim_{R_{\text{out}} \to \infty}\Phi\left(r^{\text{min}}_1\right) = \frac{r^{\text{min}}_1}{\sigma}.
\end{equation}
Assuming that $\sigma$ and hence $r^{\text{min}}_1$, are very small in comparison to $R_{\text{out}}$, this limit serves as a good approximation of the exact kinetics even when considering that a key assumption is that the timescale for $X$ exploring the neighbourhood of $E$ is small. 
\par

Returning to the bimolecular system that contains just a single enzyme molecule and $N_S$ substrate molecules, we now replicate the behaviour of the missing molecule of $X$ using the distributions from Equation (\ref{eq:AB_arrival_times}) and Equation (\ref{eq:closest_mol_CDF_infty}). That is, we mandate that $E$ only becomes reactive at times that are sampled according to Equation (\ref{eq:AB_arrival_times}) and that outside these times no reactions are possible. In addition, each time $E$ becomes reactive, we sample $r^{\text{min}}_1$ according to Equation (\ref{eq:closest_mol_CDF_infty}). The idea is that through this sampling, reactions become possible at times that are consistent with the corresponding trimolecular system. Moreover, at these times the we obtain a value for $r_1$ that is consistent with the separation between a pair of $E$ and $X$ molecules at their point of closest approach. This allows us to apply reaction conditions of the form given in Equation (\ref{eq:reactive_domain}) since we once again have two spatial degrees of freedom, namely $r^{\text{min}}_1$ and the distance to the closest molecule of $S$.\par

Notice that according to Equation (\ref{eq:closest_mol_CDF_infty}), $r^{\text{min}}_1$ is sampled from the range $[0,\sigma]$ uniformly at random. To determine whether a reaction has occurred, we then check whether $r^{\text{min}}_1$ satisfies a reaction condition of the form $r^{\text{min}}_1 \leq \sigma F(r_2)$ where $F(r_2)$ is related to $f$ from Equation (\ref{eq:reactive_domain}) by
\begin{equation}\label{eq:reaction_prob}
    F(r_2) \equiv \frac{f(r_2)}{\sigma}.
\end{equation}
The probability that this condition is satisfied is simply $\Phi_\infty(\sigma F(r_2)) = F(r_2)$. Therefore, upon $E$ becoming reactive, a reaction simply occurs with probability $F(r_2)$.\par

\begin{figure}
\centering
\includegraphics[width=0.4\textwidth]{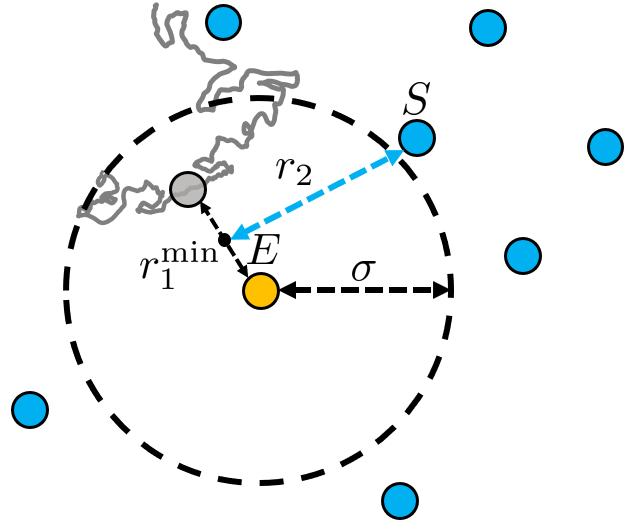}
\caption{The random sampling that mediates non-elementary bimolecular reactions can be attributed to the action of a (purely conceptual) phantom molecule, depicted by the grey partially transparent point. This phantom appears at a random point on the surface of a sphere of radius $\sigma$ surrounding a molecule of the enzyme $E$, represented by the yellow point. The arrival times of the phantom are exponentially distributed, as described by Equation (\ref{eq:AB_arrival_times}), and after appearing it explores the interior of this sphere via diffusion, as shown by the grey path. We use $r^{\text{min}}_1$ to denote the minimum radial distance attained between these molecules before the phantom diffuses away from $E$, and this value is sampled according to the cumulative distribution function in Equation (\ref{eq:closest_mol_CDF_infty}). In addition, $r_2$ denotes the distance between the centre of diffusion of these molecules at the point of closest approach and the closest molecule of the substrate $S$, depicted by the blue points.}
\label{fig:phantom}
\end{figure}

Another way to view our bimolecular simulation method is to imagine replacing the molecule of $X$ with a \textit{phantom} reactant with identical properties. The action of this phantom is made to mimic the molecule of $X$ by sampling its arrival times and its position relative to the molecule of $E$ at these times in a manner consistent with a physical molecule of $X$. As shown in Figure~\ref{fig:phantom}, from the perspective of the molecule of $E$, the phantom appears spontaneously at a radial distance of $\sigma$ away at times that are exponentially distributed according to Equation (\ref{eq:AB_arrival_times}). After appearing, the phantom diffuses---with relative diffusion constant $\hat{D}_1$---exploring the interior of a sphere of radius $\sigma$ that is centred on the molecule of $E$, before diffusing far enough away that their separation can be considered effectively infinite. We denote the radial distance between the phantom and the molecule of $E$ at the point of their closest approach $r^{\text{min}}_1$ and sample this value according to Equation (\ref{eq:closest_mol_CDF_infty}). Similarly, the distance from the centre of diffusion of the phantom and the molecule of $E$, again at their point of closest approach, to the nearest molecule of $S$ defines $r_2$. Although the phantom is a useful conceptual tool, it is crucial to remember that it is entirely fabricated and is not tracked within the simulation; rather it is used to rationalise the sampling that we ascribe to its presence.\par

Mediating reactions in this way has no effect on the steady-state reaction rate of the system since the action of the absent molecule of $X$ is entirely replicated by the phantom. In systems that contain multiple molecules of each reactant we, as is standard for particle-based models, assume that the molecules are sufficiently sparse so that molecules of the same species behave independently of one another. Under this assumption, the reaction rate is given by $K(s)$ multiplied by the respective copy numbers for the reactants besides $S$. The number of phantom molecules that one imagines also influences the reaction rate in a similar manner, but without loss of generality, we can always assume that just one immortal phantom molecule mediates the reactivity of the enzyme.\par

Under these assumptions, it is possible to construct particle-based simulations of non-elementary bimolecular reactions. For instance, recall that while the Michaelis-Menten system consists of three separate elementary reactions (see Reaction (\ref{eq:MM_network})) the rate of substrate degradation given by Equation (\ref{eq:MM_production_rate}), cannot be described by mass-action kinetics. As a consequence, particle-based simulations that use traditional reaction conditions cannot directly simulate these kinetics and instead reproduce them by explicitly simulating each of the constituent elementary reactions. This approach is inefficient because it requires resolving the very fast binding events that occur between the substrate and the enzyme. In contrast, our approach allows us to avoid explicitly modelling these binding events entirely. Instead, by applying the reaction condition in Equation (\ref{eq:MM_reaction_condition}) in the manner described here, we can directly simulate Reaction (\ref{eq:direct_MM_reaction}), which is equivalent to replicating the long-term kinetics of the system.\par

Although we have derived our approach with explicit references to Michaelis-Menten kinetics, our approach can be applied to a variety of systems, and its main limitations arise from the restrictions placed on the reaction conditions. Namely, similar to \citet{our_first_paper}, we require that the desired non-elementary kinetics be such that the corresponding reaction boundary can be determined via Equation (\ref{eq:closest_rate_old}), and that the resulting boundary is a non-negative function of $r_2$. If not all reactions are of the required form, then our method can be used along with existing techniques, and in the next section we describe how to incorporate it in particle-based simulations.\par

\section{Simulating bimolecular systems}
\label{sec:simulation_details}
Our approach can be incorporated in particle-based simulations using standard techniques, but due to the form of the reaction condition, the simulations detailed in this section resemble those presented by \citet{our_first_paper}. Although Kearney and Flegg's algorithm combines components of both time-driven approaches akin to Smolydn \cite{Andrews_2004} and event-driven approaches like eGFRD \cite{egfrd1, egfrd2} our simulations are entirely event driven. That is, the simulation progresses through time using adaptive time steps, which are determined by the next time $t_{\text{r}}$, at which a molecule becomes reactive and there is potential for a reaction to occur.\par 

The frequency of these events is dictated by the propensity the phantom has for a particular chemical species, which is characterised by the rate $K_S$, see Equation (\ref{eq:smol_bimolecular_rate}), and depends on the maximum interaction distance $\sigma$. If there are multiple molecules of a particular species present, then the propensity is given by $K_S$ multiplied by the current copy number for that species. Notice that we are assuming that the system contains just a single phantom molecule. In essence, we model the approach of the phantom to a molecule of a particular species as if this event were a bimolecular reaction. If we assume the action of the phantom is such that the reactants can always be considered well-mixed in the local vicinity of the target molecule, then this reaction can be modelled using the CME and we can use Gillespie's stochastic simulation algorithm (SSA) \cite{gillespie2007stochastic} to sample $t_{\text{r}}$. To extend this method to systems that contain multiple non-elementary bimolecular reactions, we simply associate a distinct phantom molecule, each having a potentially different propensity for approaching its target species, with each reaction. We then again leverage the SSA to determine the time of the next event, as well as which reaction this event relates to.\par

The positions of individual molecules (besides the phantom(s)) are tracked explicitly, but only need to be known when a molecules becomes reactive and the corresponding reaction condition needs to be checked. We assume that each molecule undergoes Brownian motion, and molecular positions are updated randomly each step according to \cite{Smith2019}
\begin{equation}\label{eq:brownian_dynamics}
    \pos{x}{j}(t+\Delta t) = \pos{x}{j}(t) + \sqrt{2D_j\Delta t}\pos{\xi}{j}.
\end{equation}
Here $\pos{x}{j}(t)$ denotes the position of the $j^\text{th}$ molecule at time $t$, $D_j$ is the diffusion coefficient for this molecule, $\pos{\xi}{j}$ is a three-dimensional vector of independent, normally distributed random numbers with unit variance and zero mean, and the time step $\Delta t = t_{\text{r}} - t$ is determined by the time at which the next reaction becomes possible.\par 

After updating the positions, a molecule of the reactive species is chosen uniformly at random, and $r^{\text{min}}_1$---the distance to the phantom molecule at the point of closest approach---is sampled according to the CDF in Equation (\ref{eq:closest_mol_CDF_infty}) using inversion sampling \cite{Devroye1986}. The radial distance $r_2$ is the distance from the centre of diffusion between the reactive molecule and the phantom to the closest molecule of the second reactant. The required centre of diffusion can be determined using Equation (\ref{eq:x_bari_def})
\begin{equation}
    \posbar{x}{1} = \frac{1}{D_0^{-1}+D_1^{-1}}\left(D_0^{-1}\pos{x}{0} - D_1^{-1}\pos{x}{1}\right),
\end{equation}
where the position of the phantom molecule, $\pos{x}{1}$, is sampled uniformly at random on the surface of a sphere of radius $r^{\text{min}}_1$ centred on the reactive molecule. Once $\posbar{x}{1}$ has been calculated, $r_2$ can be determined by finding the distance from this point to the closest molecule of the relevant species, and the reaction condition can be used to decide whether a reaction occurs during this event.\par

The diffusion coefficient of the phantom molecule, $D_1$, can be chosen arbitrarily because reactions involving the phantom molecule are simulated using the SSA, and the molecule itself is purely conceptual. One could imagine always scaling $\sigma$ accordingly to produce the desired reaction rate, but in practice this is unnecessary. If we assume that $D_1$ is very large (but finite) then $\posbar{x}{1} \approx \pos{x}{0}$. Under this assumption, $r_2$ is simply the distance from the reactive molecule to the closest molecule of the relevant species. In addition, it is no longer necessary to sample $r^{\text{min}}_1$, since it was only required to determine $\posbar{x}{1}$. Instead, we recall that sampling $r^{\text{min}}_1$ according to Equation (\ref{eq:closest_mol_CDF_infty}) and testing the reaction condition, is equivalent to causing a reaction to occur with probability $F(r_2)$ given by Equation (\ref{eq:reaction_prob}). Therefore, $r^{\text{min}}_1$ and $\sigma$ can be completely removed from the reaction scheme, which is not surprising since they relate to the behaviour of the phantom molecule. Our approach to simulating non-elementary bimolecular kinetics is summarised in Algorithm~\ref{alg:sim}.   

\begin{algorithm}[H]
  \begin{algorithmic}[1]
    \Function{bimolecular\_simulation}{duration, reaction\_parameters, reactants}
      \State time $\gets$ 0
      \While{time $<$ duration}
        \LineComment{Determines the time $t_r$ when a molecule next becomes reactive via Gillespie's SSA.}
        \State t\_r, reactive\_species  $\gets$ SSA(reaction\_parameters, reactants)
        \State time\_step $\gets$ t\_r - t
        \LineComment{Update the position of all molecules using Equation (\ref{eq:brownian_dynamics}).}
        \State update\_positions(reactants, time\_step)
        \State time $\gets$ time + time\_step
        \LineComment{Select a random molecule to become reactive and get its position.}
        \State reactive\_molecule, $\pos{x}{0}$ $\gets$ select\_random\_molecule(reactive\_species)
        \LineComment{Determine the distance $r_2$ to the closest molecule of the second reactant.}
        \State $r_2 \gets$ find\_closest\_molecule(reactive\_molecule, $\pos{x}{0}$)
        \LineComment{Reaction occurs with probability $F(r_2)$.}
        \State check\_for\_reaction(reactive\_species, $r_2$)
      \EndWhile
    \EndFunction
  \end{algorithmic}
  \caption{Non-elementary bimolecular kinetics simulation}
  \label{alg:sim}
\end{algorithm}

\section{Results}
\label{sec:results}
In this section, we present the results of simulations of multiple non-elementary bimolecular reactions conducted using our framework. We validate our approach in Section~\ref{subsec:MM_results} by simulating a system governed by Michaelis-Menten kinetics and comparing the simulated kinetics with theoretical predictions. Then, in Section~\ref{subsec:oscillations}, we apply our method to a more complex biological system by simulating \citeauthor{og_circadian_model}'s \cite{og_circadian_model} minimal model of circadian rhythms in \textit{Drosophila melanogaster} (fruit fly). This system allows us to demonstrate how our approach can be incorporated alongside existing particle-based techniques, and also enables our results to be compared to those of the CME, which has previously been used to study this model \cite{gonze2002}.

\subsection{Michaelis-Menten kinetics}
\label{subsec:MM_results}

To demonstrate the validity of our approach, we consider a cubic domain of volume $V$ with periodic boundary conditions, that contains $N_E$ and $N_S$ molecules of $E$ and $S$, respectively. We assume that molecules of $E$ are added to the domain at a constant rate $k_0$ and that they are removed at the same rate as the substrate is in Reaction (\ref{eq:direct_MM_reaction}). That is, we consider the reactions
\begin{equation}\label{eq:bimolecular_reaction}
    \ce{$\emptyset$ ->[k_0] E}  \quad \text{and} \quad \ce{S + E ->[{k}(s)] S},
\end{equation}
where we recall that $k\left(s\right) = k_2/(K_\mathrm{M} + s)$.\par

Notice that here the enzyme is consumed rather than the substrate. This reversal of roles is intentional and convenient, as it allows us to investigate how the simulated reaction rate varies with the substrate concentration $s$, without $s$ changing over time. By observing the steady-state enzyme concentration for various values of $s$, we are able to determine the corresponding steady-state reaction rate. In practice, this modification means that when a molecule of $E$ reacts with a molecule of $S$, we remove the enzyme from the system rather than the substrate. Otherwise, the reaction mechanics remain unchanged from those described in Section (\ref{sec:modelling_bimolecular_systems}).\par

In this section, all quantities are assumed to be dimensionless. We nondimensionalise lengths with respect to a characteristic length scale $L$ such that the dimensionless volume of the simulation domain is $V=240$. Similarly, a characteristic timescale is chosen such that the dimensionless diffusion constants for $S$ and $E$ are $D_0 = D_2 = 1$. The remaining parameters are $k_2=0.1$, $K_\text{M}=1$ and $k_0 = k_2/V$. With these choices, the steady-state distribution of $N_E$ is a Poisson distribution \cite{erban2007practical},
\begin{equation}
\label{eq:NE_distribution}
    N_E = \text{Pois}\parentheses{(}{\frac{1+\bar{s}}{\bar{s}}}{)},
\end{equation}
where
\begin{equation}
    \bar{s} = \frac{N_S}{K_\text{M}V} = \frac{N_S}{240}.
\end{equation}

At the beginning of each simulation, the initial value of $N_E$ is sampled according to Equation (\ref{eq:NE_distribution}), while the value of $N_S$ is varied so that $N_S \in\ \{24m: m=2,3,...,19\} \cup \{240n: n=2,3,...,10 \}$, which is equivalent to $\bar{s} \in\{0.1m: m=2,3,...,19\} \cup \{n: n=2,3,...,10 \}$. Notice that we have been careful to ensure that the domain contains enough molecules of $S$ for the probability density assumed by Equation (\ref{eq:closest_rate_old}) to be accurate \cite{our_first_paper}. Initially, molecules are distributed uniformly at random throughout the domain, and any molecules of $E$ that are produced also have their initial position chosen uniformly at random. The simulation is run for a dimensionless duration of $200$ units to ensure that the system has reached a steady state, at which point the value of $N_E$ is sampled.\par 

\begin{figure}[t!]
\centering
\begin{subfigure}{0.49\textwidth}
  \centering
  \includegraphics[width=1\textwidth]{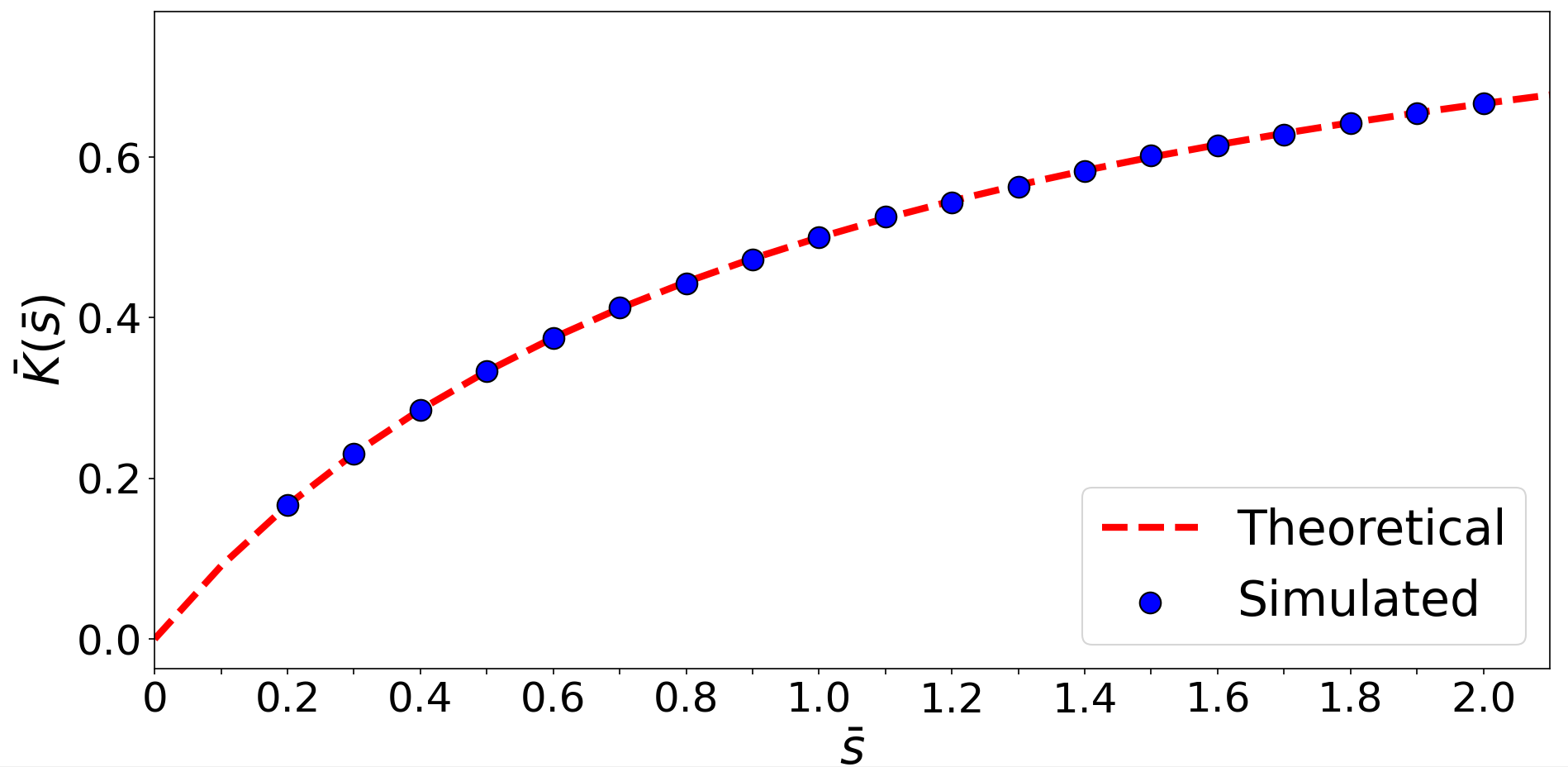}
  \caption{}
  \label{fig:rates_0.2_1}
\end{subfigure}
\begin{subfigure}{0.49\textwidth}
  \centering
  \includegraphics[width=1\textwidth]{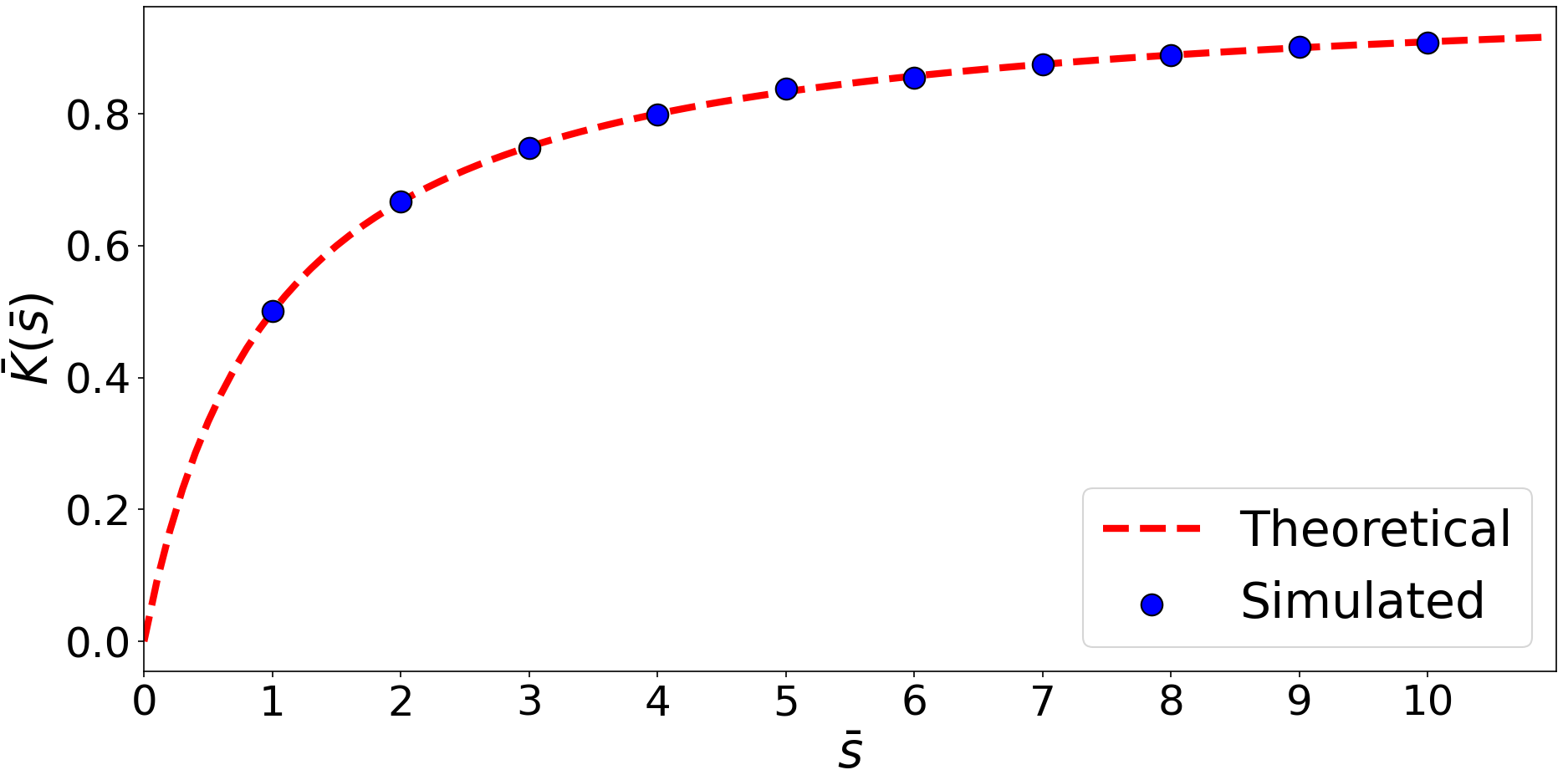}
  \caption{}
  \label{fig:rates_1_20}
\end{subfigure}\\
\begin{subfigure}{0.49\textwidth}
  \centering
  \includegraphics[width=1\textwidth]{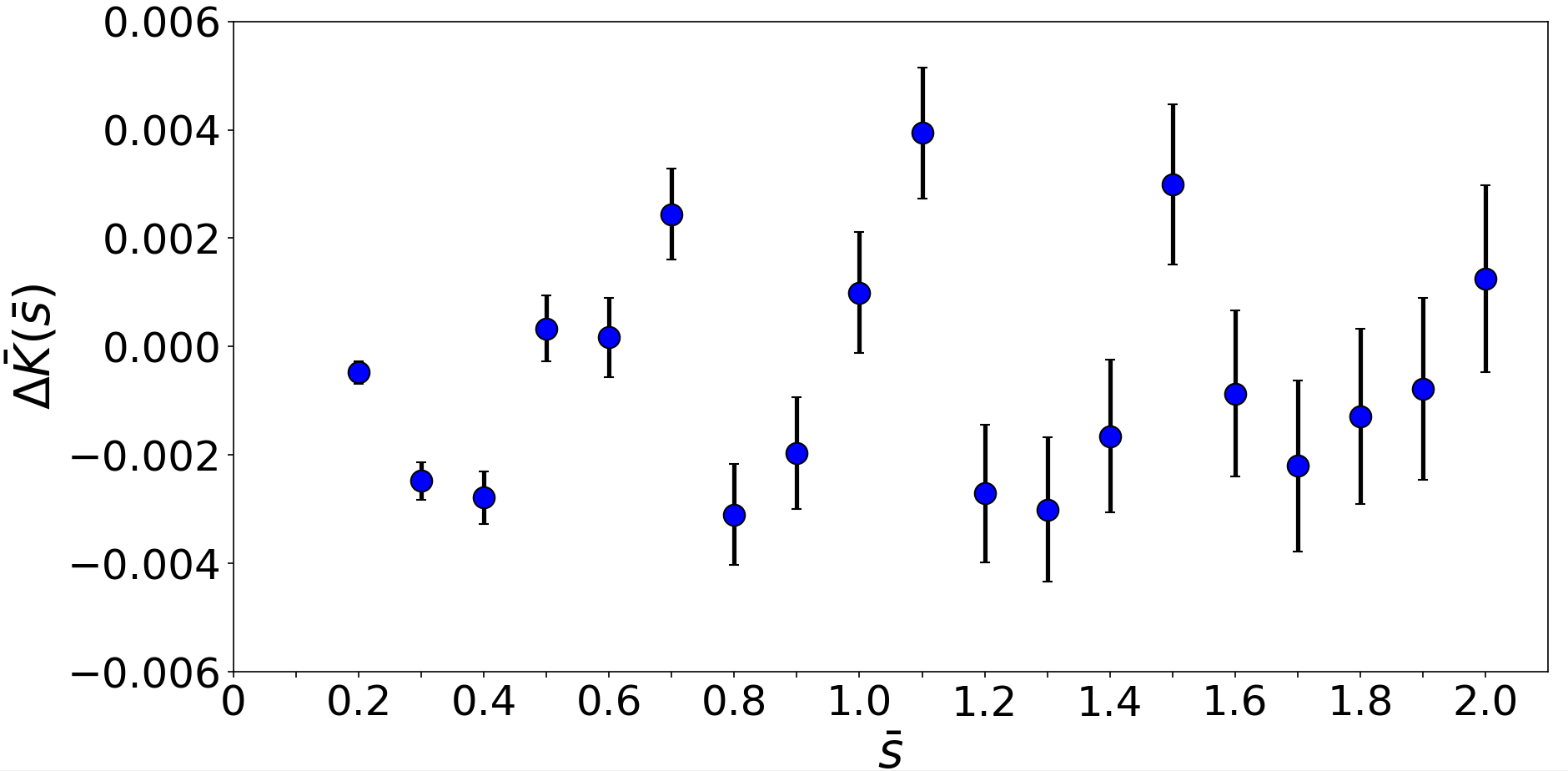}
  \caption{}
  \label{fig:errors_0.2_1}
\end{subfigure}
\begin{subfigure}{0.49\textwidth}
  \centering
  \includegraphics[width=1\textwidth]{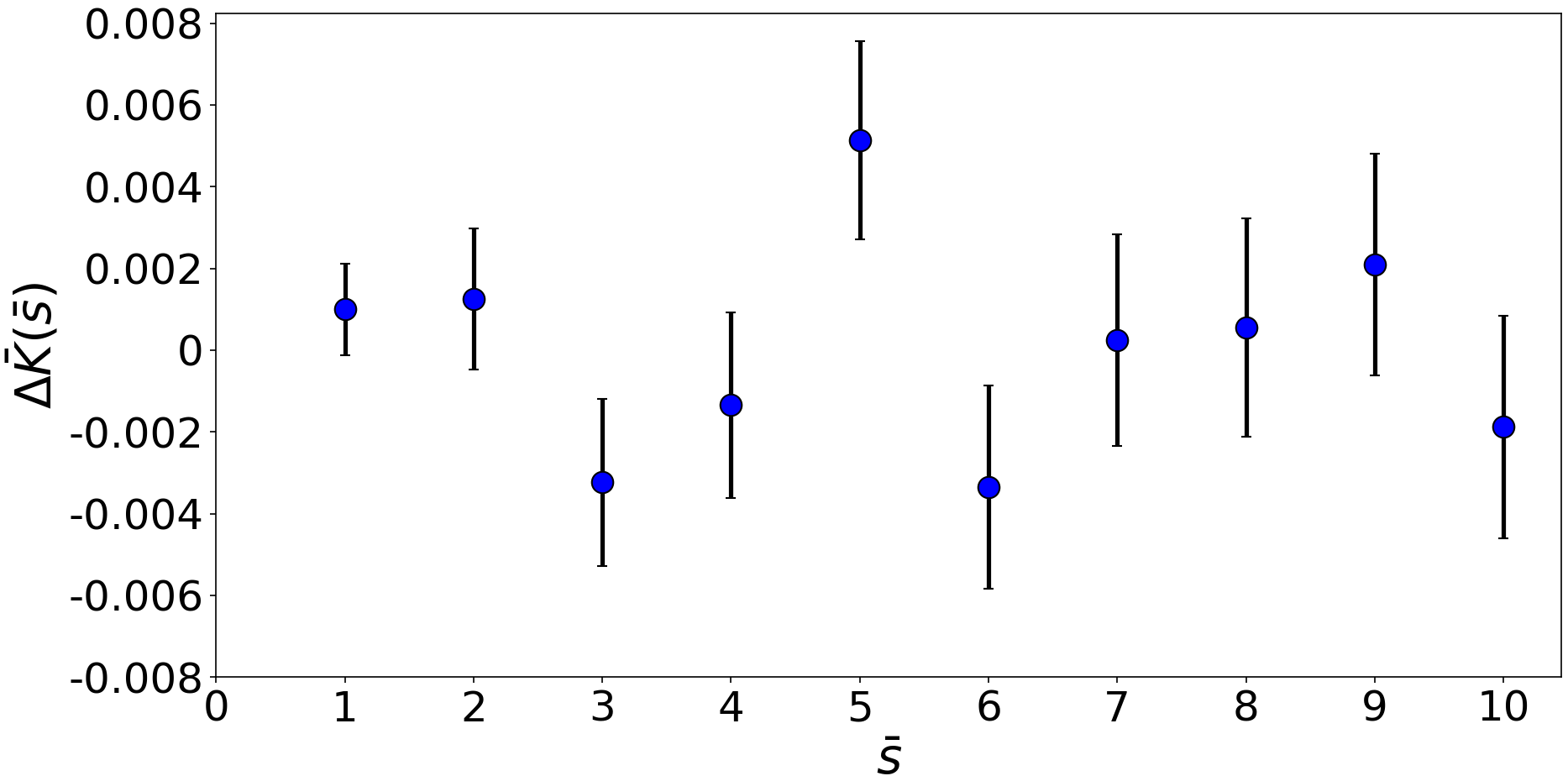}
  \caption{}
  \label{fig:errors_1_20}
\end{subfigure}
\caption{The dimensionless reaction rate for Reaction (\ref{eq:bimolecular_reaction}) for $\bar{s} \in\{0.1m: m=2,3,...,19\} \cup \{n: n=2,3,...,10 \}$. The red dashed lines in (a) and (b) show the dimensionless reaction rate $\bar{K}(\bar{s})$, see Equation (\ref{eq:expected_non_linear_rate}), for $\bar{s} \leq 2$ and $1 \leq \bar{s} \leq 10$ respectively, while the blue points denote the corresponding reaction rate $\bar{K}_S(\bar{s})$ calculated from $10^5$ particle-based simulations. Similarly the plots in (c) and (d) show the corresponding relative error $\Delta\bar{K}(\bar{s})$ defined in Equation (\ref{eq:rate_rel_error}) for the data points in (a) and (b) respectively. The black error bars in (c) and (d) denote the standard error of the mean for $\bar{K}(\bar{s})$, which is also the uncertainty in $\Delta\bar{K}(\bar{s})$. These errors are not large enough to be visible in (a) and (b).}
\label{fig:rate_test_1_20}
\end{figure}

We perform $10^5$ simulations for each value of $\bar{s}$ and then calculate the dimensionless reaction rate ($\bar{K}(\bar{s})$), which is given by the inverse of the mean of the steady-state distribution for $N_E$ in Equation (\ref{eq:NE_distribution}),
\begin{equation}
\label{eq:expected_non_linear_rate}
    \bar{K}(\bar{s}) = \frac{\bar{s}}{1+\bar{s}}.
\end{equation}
The corresponding simulated reaction rate ($\bar{K}_S(\bar{s})$) is calculated by taking the inverse of the mean of the steady-state distribution obtained for $N_E$ for each value of $\bar{s}$. These simulated rates are shown by the blue points in Figure \ref{fig:rate_test_1_20}, where they are compared to a plot of Equation (\ref{eq:expected_non_linear_rate}) shown by the red dashed line. The signed relative error 
\begin{equation}
\label{eq:rate_rel_error}
    \Delta\bar{K}(\bar{s}) = \frac{\bar{K}_S(\bar{s}) - \bar{K}(\bar{s})}{\bar{K}(\bar{s})},
\end{equation}
between the theoretical and simulated reaction rate is calculated for each value of $\bar{s}$ and shown in Figures \ref{fig:errors_0.2_1} and \ref{fig:errors_1_20}. Notice that we have divided the range of $\bar{s}$ so that Figures \ref{fig:rates_0.2_1} and \ref{fig:errors_0.2_1} show the reaction rate and associated relative error for small concentrations ($\bar{s} \leq 2$), while Figures \ref{fig:rates_1_20} and \ref{fig:errors_1_20} display the same quantities for large concentrations ($1 \leq \bar{s} \leq 10$). This separation was chosen to better highlight the non-linearity of the reaction rate near $\bar{s}=1$. Moreover, based on the findings of \citet{corrections_paper}, we expect the magnitude of the relative error to increase as $\bar{s}$ decreases.

\subsection{Circadian rhythm oscillations}
\label{subsec:oscillations}
In this section, we apply our approach to \citeauthor{og_circadian_model}'s \cite{og_circadian_model} minimal model of \textit{period} protein (PER) oscillations in \textit{Drosophila}. The period protein plays a central role in a negative feedback loop that produces oscillatory gene expression, driving circadian rhythms that regulate behaviours such as locomotor activity and the pupal emergence of adults \cite{KLARSFELD2003161,Circadian_Organization}. The circadian rhythms of \textit{Drosophila melanogaster} (fruit fly), in particular, have been extensively studied \cite{PESCHEL20111435}, owing in part to the relative simplicity of their neuronal organisation, which has now been mapped in its entirety \cite{dorkenwald2024, schlegel2024}. Of the approximately 140,000 neurons in the fruit-fly brain, a network of just 150 neurons controls circadian rhythms, making it feasible to study this phenomenon at a subcellular level \cite{10.1371/journal.pgen.1006613}. Moreover, the molecular basis of circadian rhythms in fruit flies and mammals (including humans) shows a high degree of homology (similarity due to shared ancestry) which further motivates their study \cite{TUREK2001475}.\par

Within clock neurons, PER inhibits its own transcription, resulting in a negative feedback loop that drives oscillations in its expression. In short, the transcription factors \textit{clock} (CLK) and \textit{cycle} (CYC) form a heterodimer that binds to DNA sequences in the nucleus, promoting the expression of the \textit{period} and \textit{timeless} (TIM) genes \cite{PESCHEL20111435}. This stimulates the production of PER and TIM mRNA, which is translated into the corresponding proteins in the cytoplasm. The accumulation of these proteins in the cytoplasm occurs primarily after dark because TIM stabilises PER but is itself degraded by the photoreceptor \textit{cryptochrome} \cite{hunter_ensor1996, PESCHEL2009241}. Unbound PER proteins are phosphorylated by kinase \textit{double-time} \cite{price1998, kloss1998} and subsequently degraded, although this is counteracted by the dephosphorylation caused by \textit{protein phosphatase 2A} \cite{sathyanarayanan2004}. Proteins of PER and TIM form a heterodimer that translocates into the nucleus, where it binds to the CLK/CYC dimer and inhibits the activation of the PER and TIM genes. Nuclear translocation of PER is facilitated by phosphorylation via the proteins \textit{casein kinase 2} \cite{lin2002, akten2003, Lin11175} and \textit{shaggy} \cite{Ko12664}.\par

The post-translational regulation of PER and TIM proteins induces a temporal delay between the activation of their respective genes and their subsequent repression \cite{Circadian_Organization}. Although the precise role of phosphorylation in modulating the accumulation of PER in the cytoplasm and its subsequent translocation to the nucleus remains unclear \cite{Ko12664}, the importance of this mechanism has long been recognised. As a result, it resides at the core of \citeauthor{og_circadian_model}'s \cite{og_circadian_model} model for the regulation of PER. \citeauthor{og_circadian_model} considered just five species: \textit{period} mRNA, nucleated PER, unphosphorylated, monophosphorylated and biphosphorylated cytoplasmic PER, whose concentrations---$M$, $P_\text{N}$ ,$P_0$, $P_1$ and $P_2$ respectively---are governed by the system of ODEs
\begin{subequations}
\label{eq:fly_ODEs}
\begin{align}
    \frac{dM}{dt} &= v_\text{s}\frac{K_\text{I}^n}{K_\text{I}^n + P_\text{N}^n} - v_\text{m}\frac{M}{K_\text{m} + M}\label{eq:M_ODE},\\
    \frac{dP_0}{dt} &= k_\text{s}M - V_1\frac{P_0}{K_1 + P_0} + V_2\frac{P_1}{K_2 + P_1}\label{eq:P0_ODE},\\
    \frac{dP_1}{dt} &= V_1\frac{P_0}{K_1 + P_0} - V_2\frac{P_1}{K_2 + P_1} - V_3\frac{P_1}{K_3 + P_1} + V_4\frac{P_2}{K_4 + P_2}\label{eq:P1_ODE},\\
    \frac{dP_2}{dt} &= k_\text{C}P_\text{N} - k_\text{N}P_2  + V_3\frac{P_1}{K_3 + P_1} - V_4\frac{P_2}{K_4 + P_2}  -   v_\text{d}\frac{P_2}{K_\text{d} + P_2}\label{eq:P2_ODE} \quad \text{and}\\
    \frac{dP_\text{N}}{dt} &= k_\text{N}P_2 - k_\text{C}P_\text{N}\label{eq:PN_ODE}.     
\end{align}
\end{subequations}
The first term in Equation (\ref{eq:M_ODE}) describes the transcription of $M$ in the nucleus, which is inhibited by $P_\text{N}$ according to non-elementary, Hill-type kinetics, while the second term describes the enzyme-mediated degradation of $M$ within the cytoplasm. Cytoplasmic $M$ is translated into $P_0$, which is subsequently phosphorylated to produce $P_1$. $P_1$ may revert to $P_0$ through dephosphorylation or be phosphorylated further to form $P_2$. Similarly, $P_2$ can revert to $P_1$ or it can be degraded, mimicking the action \textit{double-time} kinase. Finally, only the fully phosphorylated form ($P_2$) is shuttled to and from the nucleus, which  is captured by the first two terms in Equation (\ref{eq:P2_ODE}) and Equation (\ref{eq:PN_ODE}).\par
Even with the brief background provided here, it is clear that \citeauthor{og_circadian_model}'s model must be considered minimal, in that it overlooks many known reactants and reactions. For instance, it neglects TIM and its interactions with PER entirely. Similarly, all kinases and phosphatases that act on PER are treated implicitly, and interactions with these enzymes are captured using Michaelis-Menten kinetics. We could consider more complex models such as that of \citet{PER_TIM_model}, who extended \citeauthor{og_circadian_model}'s to incorporate TIM (although they still omit kinases and phosphatases), or their later, more comprehensive model of the mammalian circadian rhythm \cite{doi:10.1073/pnas.1132112100}. However, we choose to focus on \citeauthor{og_circadian_model}'s original model for several reasons. First, its constituent reactions are either unimolecular or of a non-elementary form that is amenable to our simulation framework. While such reactions are convenient in the current context, this model is not unique in this regard. Many models \cite{doi:10.1073/pnas.87.4.1461, NFKB, MENGEL2010656} rely on non-elementary kinetics to incorporate the action of reactants (typically enzymes) that are not of direct interest to the modeller. This approach reduces the number of explicitly reactants and reactions, decreasing both conceptual and analytic complexity and making the model more amenable to computational simulation. The ability to directly simulate common non-elementary kinetics, without resolving the underlying enzymatic binding events, is the primary advantage of our framework.\par

Another advantage of \citeauthor{og_circadian_model}'s model is that it includes only five reactants, all of which occur at concentrations low enough to allow particle-based simulation of a realistically sized system. Moreover, the model implicitly incorporates reactant transport in that PER is modelled as moving between the cytoplasm and nucleus via first-order reactions, while the translocation of mRNA from the nucleus to the cytoplasm is implicitly captured within its production. This motion is critical to the system's stability and can be modelled explicitly in a particle-based model. Finally, it is important to emphasise that our intention in this article is not to study circadian rhythms in fruit flies, but rather to demonstrate the utility of our framework in a biologically relevant setting.\par

The original parameters for the model were only ascribed a `tentative' scale and chosen to yield a period of approximately $24$ hours. Therefore, we instead adopt the revised values of \citet{PER_TIM_model} who reduced the characteristic concentrations from $\mu$M to nM, which is consistent with  a related model of circadian rhythms in mammals \cite{doi:10.1073/pnas.1132112100}, and yields counts of \textit{period} mRNA consistent with those found by \citet{10.1371/journal.pgen.1006613}. The parameters: $V_1 = V_3 = 8\text{nMh}^{-1}$, $V_2 = V_4 = 1\text{nMh}^{-1}$, $K_1 = K_2 = K_3 = K_4 = 2\text{nM}$, $K_\text{m} = K_\text{d} = 0.2\text{nM}$ and $k_\text{s} = 0.9\text{h}^{-1}$ match \citeauthor{PER_TIM_model}; while the parameters: $v_\text{s} = 1.4\text{nMh}^{-1}$, $v_\text{d} = 3\text{nMh}^{-1}$, $K_\text{I} = 2\text{nM}$, $k_\text{N} = 0.67\text{h}^{-1}$ and $k_\text{C} = 0.3\text{h}^{-1}$ have been altered because we set $n = 2$ rather than $n = 4$ since the latter does not yield a valid reaction boundary. Moreover, because we only model PER and neglect the complex it forms with TIM, the parameters $k_\text{N}$ and $k_\text{C}$ are not equivalent to those in \citeauthor{PER_TIM_model}'s model and are instead chosen to yield qualitatively similar behaviour.\par

\begin{figure}
    \centering
    \includegraphics[width=\linewidth]{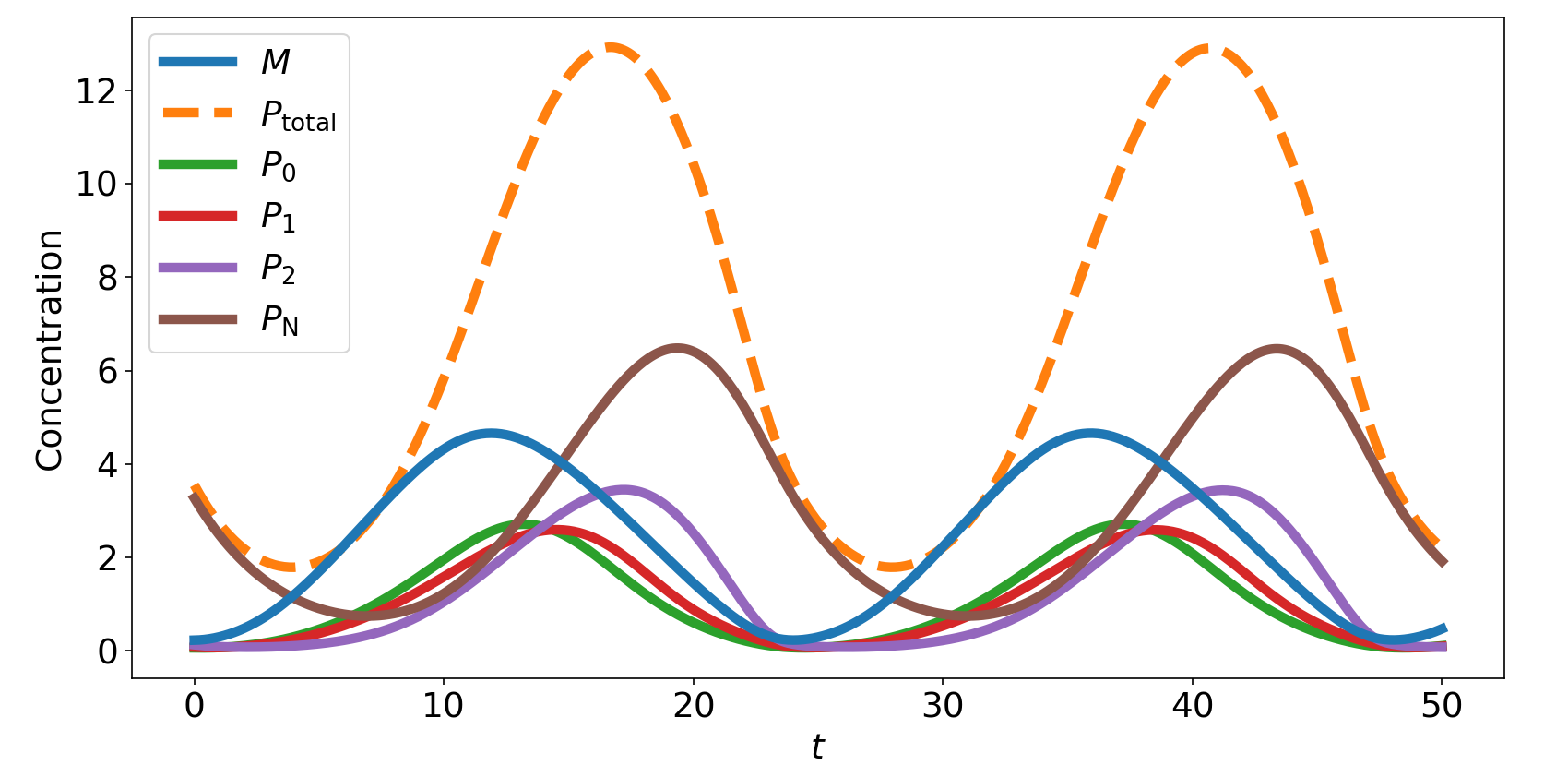}
    \caption{The solution to the ODE model in Equation (\ref{eq:fly_ODEs}). The concentration for each reactant as well as the total amount of PER $P_\text{total} = P_0+P_1+P_2+P_\text{N}$ is displayed in nM, while the time $t$ is given in hours. The parameters for the model are described in the main text and were chosen to yield behaviour qualitatively similar to that reported by \citet{og_circadian_model} and \citet{PER_TIM_model}. The summary statistics for the solution are provided in Table \ref{tab:fly_statistics}.}
    \label{fig:fly_ODE_solution}
\end{figure}

The solution to the ODE system in Equation (\ref{eq:fly_ODEs}) for our chosen parameters exhibits sustained oscillations with a period of $24.04$h as shown in Figure~\ref{fig:fly_ODE_solution}. Two full periods appear in the displayed interval $t \in [0,50]$ and $M$, $P_0$, $P_1$, $P_2$ and $P_\text{N}$ are shown by the solid blue, green, red, purple, and brown lines, respectively. The total amount of PER $P_\text{total} = P_0+P_1+P_2+P_\text{N}$, is shown by the dashed orange line. All concentrations are given in nM, time $t$ is in hours, and the initial values were determined by the point on the limit cycle where $M$ reaches its minimum. The average concentration and the amplitude of oscillation for each species is displayed in Table \ref{tab:fly_statistics}. \par 

To construct a particle-based simulation of the system in Equation (\ref{eq:fly_ODEs}) we first note that all reactions (except one) are either unimolecular elementary reactions, which can be modelled as Poissonian processes that are independent of molecular diffusion \cite{Andrews_2004,egfrd2}, or obey Michaelis-Menten kinetics. To simulate the reactions that obey Michaelis-Menten kinetics, we use the reaction condition from Equation (\ref{eq:MM_reaction_condition}), which corresponds to the reaction probability
\begin{equation}\label{eq:MM_reaction_prob}
    F_\text{MM}(r_2) = \mathrm{exp}\left(\frac{-4\pi  \Gamma r_2^3}{3}\right),
\end{equation}
where $\Gamma$ is replaced with the appropriate Michaelis constant for each reaction. The only remaining reaction is the production of $M$ (the first term in Equation (\ref{eq:M_ODE})), which is inhibited by $P_\text{N}$ according to Hill-type kinetics. For $n=2$, the corresponding reaction probability can be found by inverting Equation (\ref{eq:closest_rate_old}) and dividing by the corresponding reaction radius $2v_\text{s}/(4\pi \hat{D}_1)$, which yields
\begin{equation}\label{eq:Hill_reaction_prob}
    F_{\text{H}2}(r_2) = \mathrm{sin}^2\left(\frac{4\pi  K_\text{I}r_2^3}{6}\right).
\end{equation}\par

Notice that all of the non-elementary reactions are actually unimolecular rather than bimolecular. This is because the model assumes all enzyme concentrations are constant and well-mixed. On the surface this poses an issue for our method since we require the position of a specific enzyme in order to determine $r_2$, and hence test for a reaction whenever the enzyme becomes reactive. However, because the enzymes are well-mixed, and their concentrations have been incorporated into the associated rate constant for each reaction, it is sufficient to sample the position of the required enzyme uniformly at random within the domain whenever a reaction condition needs to be tested. If the enzymes are not well-mixed on the scale of the domain then we could track them explicitly, in which case our original bimolecular approach applies.\par

We present two variants of our particle-based simulation of \citeauthor{og_circadian_model}'s model, which differ in their treatment of the transfer of PER between the cytoplasm and nucleus, and in the boundary conditions they impose. In the first and simpler simulation, we do not explicitly distinguish the nucleus from the cytoplasm, so the exchange between $P_2$ and $P_\text{N}$ is governed simply by the elementary unimolecular reactions in Equation (\ref{eq:P2_ODE}) and Equation (\ref{eq:PN_ODE}). That is, $P_2$ can be converted to $P_\text{N}$, and vice versa, at any time, regardless of its position within the domain. The simulation domain is cubic, and we impose periodic boundary conditions in an attempt to study the reaction mechanisms independently of spatial effects. We will henceforth refer to this variant as the calibration simulation. \par

\begin{figure}
    \centering
    \includegraphics[width=0.6\linewidth]{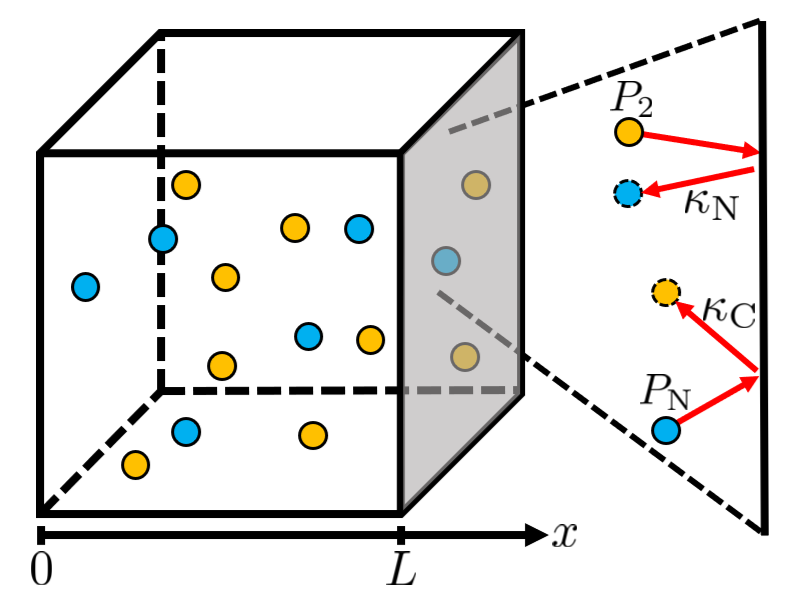}
    \caption{Simulation domain for the spatial particle-based simulation. The domain is a cube of volume $V=10^{-16}$m$^3$ (side length $L = V^{\frac{1}{3}}$) with reflective boundary conditions imposed on all faces. The planar boundary at $x=L$ mimics the nuclear membrane in that molecules of $P_2$ (the yellow points) and $P_\text{N}$ (the blue points) are converted into one another upon reflection, with the probability of conversion determined by the permeabilities $\kappa_\text{N}$ and $\kappa_\text{C}$ respectively. This process is shown in the inset on the right. All other reactants not shown here undergo normal reflection at this boundary.}
    \label{fig:simulation_domain}
\end{figure}

The second variant, referred to henceforth as the spatial simulation, is more realistic because it confines the molecules by imposing reflective boundary conditions at the edges of the domain and explicitly models the transfer of PER between the cytoplasm and nucleus in space. A single face of the cubic simulation domain is selected to act as the nuclear membrane as shown in Figure~\ref{fig:simulation_domain}. The reactants $M$, $P_0$, and $P_1$ reflect off the membrane as they would any of the other boundaries, since they are assumed to be confined to the cytoplasm. In contrast, any molecule of $P_2$ ($P_\text{N}$) incident on the membrane has a probability of being transmitted through the membrane into the nucleus (cytoplasm) to become a molecule of $P_\text{N}$ ($P_2$).\par

The probability of transmission in steady-state given a known time step and diffusion coefficient has been calculated by \citet{Andrews_2009}, but their expression requires the membrane permeability. We use $\kappa_\text{N}$ to denote the permeability of the membrane when moving from the cytoplasm to the nucleus and $\kappa_\text{C}$ to denote the permeability when moving from the nucleus to the cytoplasm. To relate $\kappa_\text{N}$ to the corresponding transfer rate $k_\text{N}$, we note that $k_\text{N}$ is the total diffusive flux across the membrane into the nucleus of the steady-state probability density ($Q(\boldsymbol{x})$) to find a molecule of $P_2$ at position $\boldsymbol{x}$. Therefore, by  the definition of the permeability, we must have
\begin{equation}\label{eq:permability_full}
    k_\text{N} = \kappa_\text{N}\int Q(\boldsymbol{x})dS_\text{M},
\end{equation}
where $dS_\text{M}$ is an infinitesimal surface-area element on the membrane and we have assumed that $\kappa_\text{N}$ is constant over the membrane. For reversible transmission, the steady-state probability density is well mixed, although discontinuous across the membrane \cite{Andrews_2009}, and Equation (\ref{eq:permability_full}) simplifies to
\begin{equation}\label{eq:permability_simple}
    k_\text{N} = \frac{\kappa_\text{N}A_\text{M}}{V_\text{C}}, 
\end{equation}
where $A_\text{M}$ is the surface area of the nuclear membrane and $V_\text{C}$ is the volume of the cytoplasm.\par 

An expression equivalent to Equation (\ref{eq:permability_simple}) is used to relate $k_\text{C}$ to $\kappa_\text{C}$ where we note that in this case $V_\text{C}$ is replaced by the volume of the nucleus. However, we do not explicitly separate the nucleus from the cytoplasm. Instead, whenever a molecule of $P_2$ or $P_\text{N}$ is reflected from the membrane, it is relabelled to $P_\text{N}$ or $P_2$, respectively, with the probabilities given by \citet{Andrews_2009} (see Equation ($47$) and Equation ($48$) therein). This is entirely equivalent to treating the nucleus as a distinct spatial region since $P_\text{N}$ does not interact with any other reactants, but is more convenient computationally. We are implicitly assuming that cytoplasm and nucleus are of equal size, which while not physical, is accounted for in the transmission probability for $P_\text{N}$ to leave the nucleus (switch to a molecule of $P_2$). Finally, because Algorithm~\ref{alg:sim} is event driven, the time step can be too large to accurately resolve collisions with the membrane. To rectify this, we introduce a minimum time step $\Delta t_\text{min}$, which is used to update the molecular positions each iteration until the time step calculated on line $6$ of the algorithm is smaller than or equal to $\Delta t_\text{min}$. This allows us to control the resolution with which we resolve the transmission of $P_2$ and $P_\text{N}$ and produces an algorithm akin to the hybrid methodology used previously for the simulation of non-elementary trimolecular reactions \cite{our_first_paper}.\par 

\begin{figure}[!ht]
    \centering
    \includegraphics[width=\linewidth]{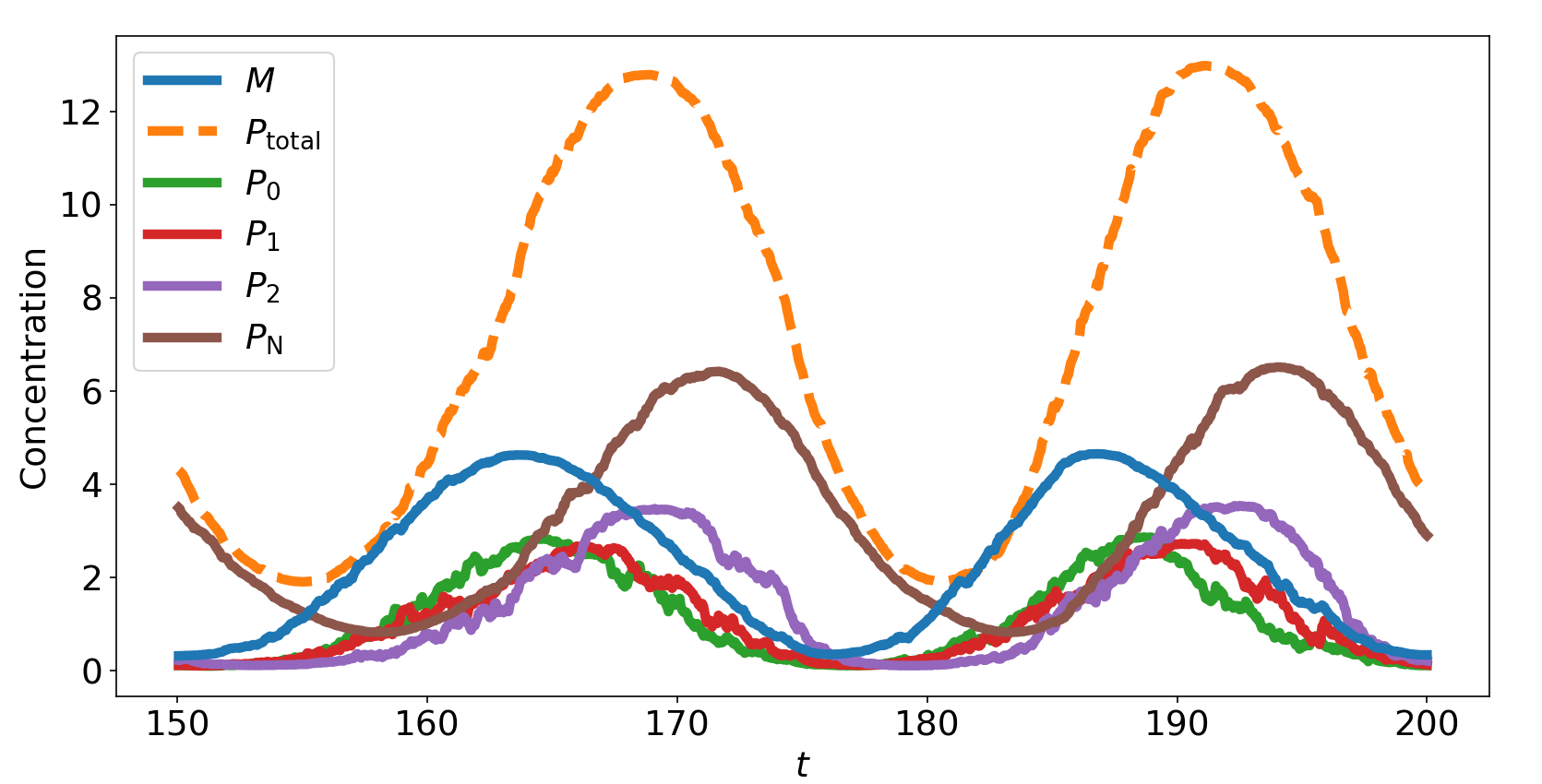}
    \caption{A phase aligned reconstruction of the average reactant concentrations from $500$ replicates of the SSA  applied to the system described by Equation (\ref{eq:fly_ODEs}). The concentration for each reactant as well as the total amount of PER $P_\text{total} = P_0+P_1+P_2+P_\text{N}$ is displayed in nM, while the time $t$ is given in hours. The interval $t\in[150,200]$ is chosen to ensure the system has reached a stable limit cycle while avoiding artefacts in the reconstruction present at the extremes of the simulated interval. Initial concentrations are equivalent to those used for the numerical solution to the ODE shown in Figure \ref{fig:fly_ODE_solution} and the simulation domain is a cube of volume $10^{-16}\text{m}^3$. The corresponding summary statistics are provided in Table \ref{tab:fly_statistics}.}
    \label{fig:fly_gillespie_solution}
\end{figure}

\begin{figure}[!ht]
    \centering
    \includegraphics[width=\linewidth]{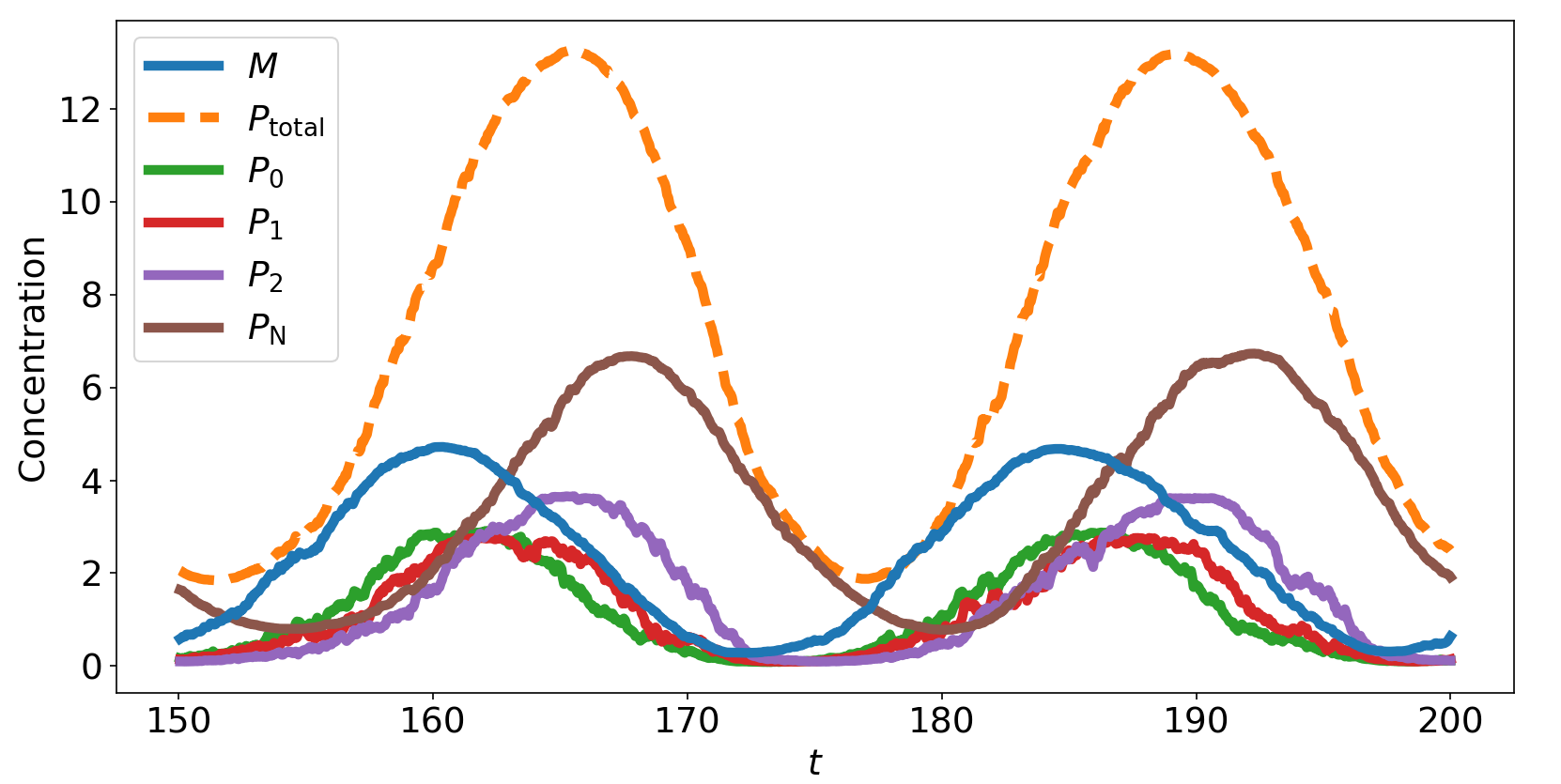}
    \caption{A phase aligned reconstruction of the average reactant concentrations from $500$ replicates of the calibration particle-based simulation applied to the system described by Equation (\ref{eq:fly_ODEs}). The concentration for each reactant as well as the total amount of PER $P_\text{total} = P_0+P_1+P_2+P_\text{N}$ is displayed in nM, while the time $t$ is given in hours. The interval $t\in[150,200]$ is chosen to ensure the system has reached a stable limit cycle while avoiding artefacts in the reconstruction present at the extremes of the simulated interval. Initial concentrations are equivalent to those used for the numerical solution to the ODE shown in Figure \ref{fig:fly_ODE_solution}. The simulation domain is a cube of volume $10^{-16}\text{m}^3$ with periodic boundary conditions and the diffusion coefficient for all species is $10^{-9}\text{m}^2\text{h}^{-1}$. The corresponding summary statistics are provided in Table \ref{tab:fly_statistics}.}
    \label{fig:fly_calibration_solution}
\end{figure}

\begin{figure}[!ht]
    \centering
    \includegraphics[width=\linewidth]{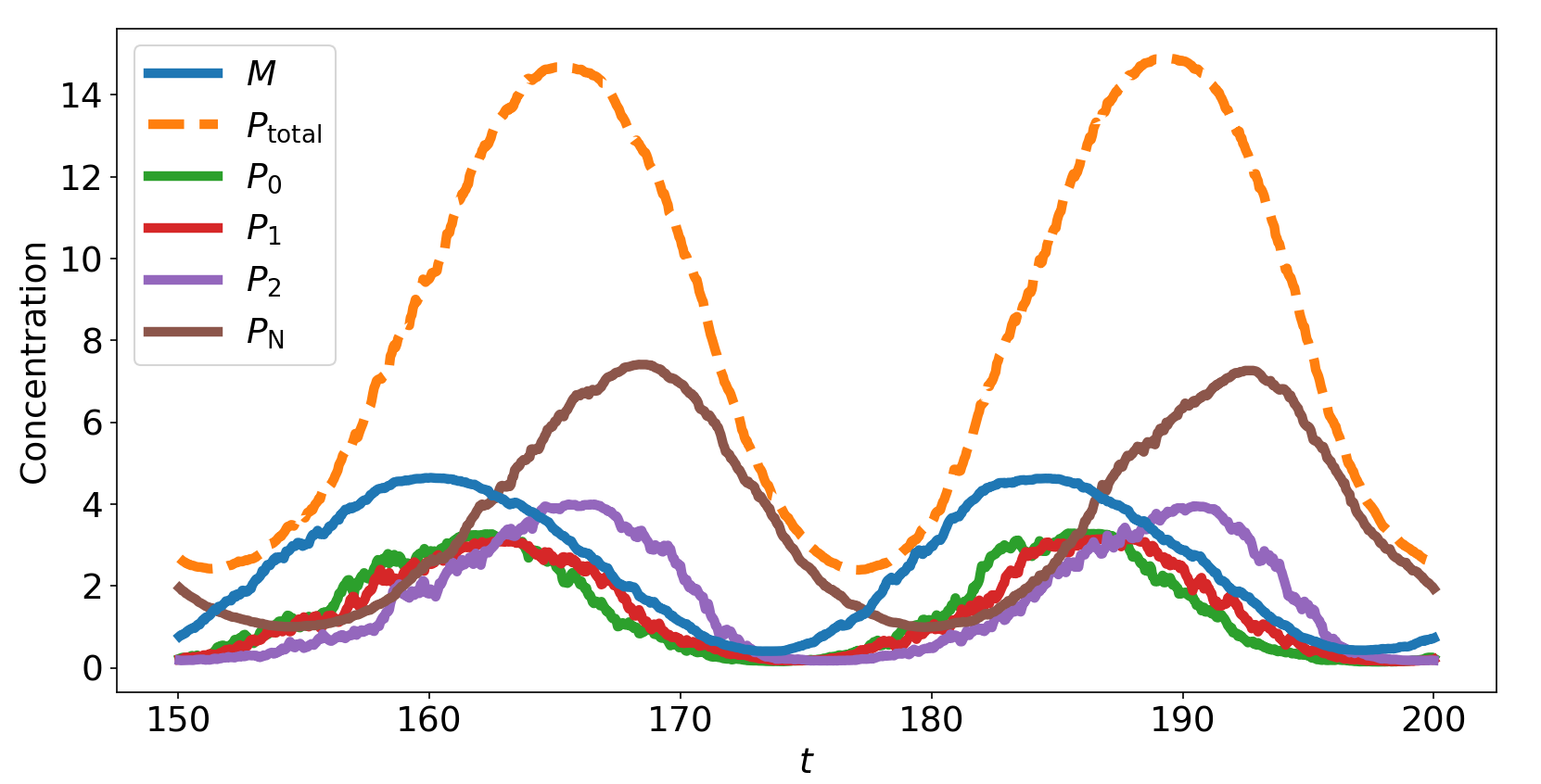}
    \caption{A phase aligned reconstruction of the average reactant concentrations from $500$ replicates of the spatial particle-based simulation applied to the system described by Equation (\ref{eq:fly_ODEs}). The concentration for each reactant as well as the total amount of PER $P_\text{total} = P_0+P_1+P_2+P_\text{N}$ is displayed in nM, while the time $t$ is given in hours. The interval $t\in[150,200]$ is chosen to ensure the system has reached a stable limit cycle while avoiding artefacts in the reconstruction present at the extremes of the simulated interval. Initial concentrations are equivalent to those used for the numerical solution to the ODE shown in Figure \ref{fig:fly_ODE_solution}. The simulation domain is a cube of volume $10^{-16}\text{m}^3$ with reflective boundary conditions and the diffusion coefficient for all species is $10^{-9}\text{m}^2\text{h}^{-1}$. In addition, a single face of the domain is chosen to act as the nuclear membrane that converts $P_2$ to $P_\text{N}$ and vice versa as shown in Figure \ref{fig:simulation_domain}. The corresponding summary statistics are provided in Table \ref{tab:fly_statistics}.}
    \label{fig:fly_spatial_solution}
\end{figure}

Since our particle-based simulations introduce stochasticity to a deterministic model, we also conduct a stochastic simulation of the system using Gillespie's SSA for the sake of comparison. The simulation domain is taken to be a cube of volume $10^{-16}$m$^3$ for all simulations which is consistent with the median neuron size in fruit flies \cite{dorkenwald2024}. Similarly, the diffusion coefficient for all species is $10^{-9}\text{m}^2\text{h}^{-1}$ for both particle-based simulations, which is consistent with the diffusion rates for both protein \cite{gregor2007} and mRNA molecules \cite{DILAO2010847} in the cytoplasm of fruit fly cells. The initial populations for each simulation are the same as for the solution to the ODE system presented in Figure~\ref{fig:fly_ODE_solution} and $300$h of evolution is simulated, although we discard the first $100$h to ensure the system has reached a stable limit-cycle. Figures \ref{fig:fly_gillespie_solution}, \ref{fig:fly_calibration_solution} and \ref{fig:fly_spatial_solution} display the evolution of the system for the SSA simulation, the calibration simulation and the spatial simulation, respectively. Each figure depicts a phase aligned reconstruction of the average concentrations from $500$ replicates and the interval $t \in [150, 200]$ is chosen to account for the discarded transient and to avoid artefacts in the reconstruction present near the extremes of the simulated interval. For each simulation, the period, average concentrations, and average oscillation amplitudes of each species are compared to the corresponding values from the ODE solution in Table~\ref{tab:fly_statistics}.\par

\begin{table}[h]
\centering
\setlength{\tabcolsep}{8pt} 
\renewcommand{\arraystretch}{1.25} 
\begin{tabular}{|l|c|c|c|c|}
\hline
\textbf{Quantity} & \textbf{ODE} & \textbf{SSA} & \textbf{calibration} & \textbf{spatial}\\
\hline
\textbf{Period ($M$)} & $24.04$ & $24.29 \pm 0.06$& $24.31 \pm 0.06$ & $25.28 \pm 0.05$\\
\hline
\textbf{Average ($M$)} & $2.399$ & $2.429 \pm 0.002$ & $2.428 \pm 0.002$ & $2.464 \pm 0.002$ \\
\hline
\textbf{Average ($P_0$)} & $1.144$ & $1.165 \pm 0.002$ & $1.171 \pm 0.002$ & $1.384 \pm 0.002$ \\
\hline
\textbf{Average ($P_1$)} & $1.140$ & $1.157 \pm 0.002$ & $1.162 \pm 0.002$ & $1.373 \pm 0.002$ \\
\hline
\textbf{Average ($P_2$)} & $1.401$ & $1.410 \pm 0.004$ & $1.428 \pm 0.004$ & $1.678 \pm 0.004$ \\
\hline
\textbf{Average ($P_\text{N}$)} & $3.129$ & $3.135 \pm 0.009$ & $3.178 \pm 0.009$ & $3.664 \pm 0.009$ \\
\hline
\textbf{Average ($P_\text{total}$)} & $6.81$ & $6.87 \pm 0.02$ & $6.94 \pm 0.01$ & $8.10 \pm 0.02$ \\
\hline
\textbf{Amplitude ($M$)} & $4.42$ & $4.45 \pm 0.02$ & $4.54 \pm 0.02$ & $4.38 \pm 0.02$ \\
\hline
\textbf{Amplitude ($P_0$)} & $2.65$ & $3.07 \pm 0.02$ & $3.11 \pm 0.02$ & $3.46 \pm 0.02$ \\
\hline
\textbf{Amplitude ($P_1$)} & $2.52$ & $2.96 \pm 0.02$ & $3.00 \pm 0.02$ & $3.30 \pm 0.02$ \\
\hline
\textbf{Amplitude ($P_2$)} & $3.36$ & $3.78 \pm 0.03$ & $3.85 \pm 0.03$ & $4.18 \pm 0.03$ \\
\hline
\textbf{Amplitude ($P_\text{N}$)} & $5.72$ & $5.89 \pm 0.06$ & $6.06 \pm 0.06$ & $6.58 \pm 0.06$ \\
\hline
\textbf{Amplitude ($P_\text{total}$)} & $11.10$ & $11.31 \pm 0.09$ & $11.59 \pm 0.09$ & $12.7 \pm 0.1$ \\
\hline
\end{tabular}
\caption{Summary statistics for the SSA, calibration, and spatial simulations, alongside deterministic values from the solution to the ODE system in Equation (\ref{eq:fly_ODEs}). The period is expressed in hours, while amplitudes and average concentrations are reported in nM. For the three stochastic simulations, all statistics are calculated by averaging over $500$ replicates, each containing $200$h of evolution; the first $100$h of the original $300$h is discarded to ensure the system has reached a stable limit-cycle. All uncertainties represent the standard error of the mean and are shown to one significant figure. Comparative values for the ODE system are derived from the numerical solution, a portion of which is shown in Figure \ref{fig:fly_ODE_solution}. The species each row refers to is indicated in parentheses.}
\label{tab:fly_statistics}
\end{table}

\section{Discussion}
\label{sec:discussion}
Our results in Section~\ref{subsec:MM_results} demonstrate good agreement between the simulated reaction rate obtained from particle-based simulations of Reaction (\ref{eq:bimolecular_reaction}) and the theoretical rate given in Equation (\ref{eq:expected_non_linear_rate}). The relative error between these quantities, shown in Figures~\ref{fig:errors_0.2_1} and \ref{fig:errors_1_20}, does not show any discernable trend. However, with greater precision, we would expect it to decrease as $\bar{s}$ increases, since this behaviour has been demonstrated by \citet{corrections_paper} who studied this error using finite element methods. Specifically, \citeauthor{corrections_paper} showed that the discrepancy between the simulated rate and the theoretical rate arises because the reaction boundary determined by Equation (\ref{eq:closest_rate_old}) only approximates the true boundary. To make this more precise, consider an arbitrary reaction boundary $\partial \Omega$, which corresponds to the reaction rate $K(s)$. That is, $K(s)$ is equal to the total flux of the steady-state probability density $P(\pos{\eta}{})$ over $\partial \Omega$. Equation (\ref{eq:closest_rate_old}) neglects a small correction to the diffusive flux of $P(\pos{\eta}{})$ in the $\pos{\eta}{1}$ direction, while ignoring the flux of $P(\pos{\eta}{})$, which consists of both a diffusive and advective component, across the boundary in the $\pos{\eta}{2}$ direction. Therefore, the reaction rate $K_1(s)$, only approximates the true reaction rate $K(s)$ associated with $\partial \Omega$. Moreover, if $K(s)$ is known, then using Equation (\ref{eq:closest_rate_old}) to determine the corresponding reaction boundary will only yield an approximation of $\partial \Omega$. The accuracy of this approximation increases with concentration because $P(\pos{\eta}{})$, and hence its flux, becomes more localised as the concentration increases, making the diffusive flux in the $\pos{\eta}{1}$ direction a better approximation of the total flux.\par

\citet{corrections_paper} demonstrate how to correct the errors introduced by using Equation (\ref{eq:closest_rate_old}) to determine the reaction boundary. Their approach retains the functional form of the reaction boundary, but modifies the parameters that define it. In this case, this would involve minimising the relative error $\Delta\bar{K}(\bar{s})$ by altering $\Gamma$ and the rate at which the phantom molecule approaches molecules of $E$. However, for our chosen parameters, the relative error is already very small, and resolving such corrections would require an impractically large number of simulation replicates.  Moreover, in our simulations, molecules of $S$ remain stationary during the time that a molecule of $E$ is reactive, effectively eliminating the flux across the boundary in the $\pos{\eta}{2}$ direction. Although this is not strictly physically, it helps mitigate the error that arises when using Equation (\ref{eq:closest_rate_old}) since it also neglects this flux. As a result, we expect the simulated reaction rate to match $K_1(s)$ more closely than the work of \citet{corrections_paper} would suggest.\par

Although our approach is based on the method developed by \citet{our_first_paper}, it is more accurate and significantly easier to implement. Algorithm~\ref{alg:sim} is event driven and molecular positions are known exactly at the time of each event. In contrast, \citeauthor{our_first_paper} were forced to adopt a hybrid scheme combining time- and event-driven components, due to the difficultly of extending fully event-driven approaches such as GFRD \cite{gfrd1, gfrd2} or eGFRD \cite{egfrd1,egfrd2} to a trimolecular system. The primary obstacle being the lack of closed-form Green's function for the three-body problem. Therefore, in their trimolecular framework, where $X$ is explicitly tracked, molecules deemed close enough to potentially react have their positions updated using finite time steps according to Equation (\ref{eq:brownian_dynamics}). At each step, the relative proximity of the molecules is calculated and the reaction condition is tested. However, updating molecular positions in this manner inherently limits spatial resolution, making it possible to miss reaction events since the trajectory of the molecules cannot be resolved between time steps. This issue can be corrected for in bimolecular systems \cite{Andrews_2004}, but such corrections do not transfer straightforwardly to trimolecular systems because the reaction radius is a function of the molecules' relative proximity. Our method completely avoids these complications because the distance to closest molecule of $S$ is calculated at the exact moment a molecule of $E$ becomes reactive.\par

The assumptions underlying our approach are similar to those identified by \citet{our_first_paper}. In particular, we assume that:
\begin{enumerate}
    \item The molecules of the second reactant $S$, the closest of which defines $r_2$, must be well-mixed on length scales comparable to the typical intramolecular separation. This is required so that, in the vicinity of a reactive molecule of the first reactant $E$, the probability density to find the closest molecule of $S$ at a radial distance $r_2$, can be approximated by 
    \begin{equation}\label{eq:boundary_free_PD}
        P(r_2) = 4\pi r^2_2s\text{exp}\left(\frac{-4\pi s r_2^3}{3}\right)
    \end{equation}
     as assumed by Equation (\ref{eq:closest_rate_old}).

    \item The system can be reduced to a single reaction, or a series of reactions, similar in form to Reaction (\ref{eq:bimolecular_reaction}). That is, the reduced reaction(s) must be bimolecular, and the reaction rate(s) must be of the form $K(s)$, where $K(s)$ is a non-elementary reaction rate that depends on the concentration $s$ for each molecule of $E$.
    
    \item The inverse Laplace transform of the reaction rate $K(s)/s$ must exist so that Equation (\ref{eq:closest_rate_old}) can be inverted to determine the function $f$ that defines the reaction boundary via Equation (\ref{eq:reactive_domain}). Furthermore, the resulting function $f$ must be non-negative for all values of $r_2$ since it denotes the distance at which molecules interact and defines the probability of reaction $F(r_2)$ through Equation (\ref{eq:reaction_prob}).
\end{enumerate}\par

Assumption~1 arises from the derivation of Equation (\ref{eq:closest_rate_old}) where we explicitly consider a well-mixed concentration of $S$ molecules. Under this assumption, Equation (\ref{eq:boundary_free_PD}) is exact in the absence of reactions---that is, it is the boundary-free solution---and serves as a good approximation to the probability density in the vicinity of the reaction boundary, as discussed in Section~\ref{sec:Review}. Because the exact probability density near the boundary is difficult to determine analytically \cite{corrections_paper}, Equation (\ref{eq:boundary_free_PD}) is used to approximate this density in Equation (\ref{eq:closest_rate_old}). Therefore, this approximate density must be valid in the vicinity of any reactive molecule of $E$ if the simulated reaction rate is to match the expected reaction rate $K_1(s)$. The $99^\text{th}$ percentile for $r_2$, assuming the probability density in Equation (\ref{eq:boundary_free_PD}), is $~s^{-\frac{1}{3}}$ and provides a practical estimate of the scale on which we require $S$ to be well mixed.\par

Assumption~2 does not impose any significant restrictions because our approach can be used in tandem with other particle-based simulation methods, be they time- or event-driven, and simply increases the variety of reaction rates that can be simulated directly. In Section~\ref{subsec:oscillations} we demonstrate this by simulating \citeauthor{og_circadian_model}'s \cite{og_circadian_model} model, which---like many biochemical systems---involves a combination of elementary and non-elementary kinetics, with the latter resulting from applying the QSSA to an underlying set of elementary reactions. The usual approach is to revert to the full system and explicitly simulate each elementary reaction. In contrast, our method, when used in combination with standard techniques, enables direct simulation of the reduced system. This reduces the number of reactions that need to be simulated, but more importantly allows fast reactions, whose action is incorporated by the QSSA, to be avoided entirely, improving overall simulation efficiency.\par

Our results, summarised in Table~\ref{tab:fly_statistics}, indicate that incorporating stochasticity in \citeauthor{og_circadian_model}'s model slightly increases both the period and amplitude of oscillations, with the values obtained from all of the stochastic simulations exceeding those from the ODE solution. The only exception is the average amplitude of the oscillations in $M$ from the spatial simulation, though even in this case, the average concentration in this case still exceeds the ODE value. These findings are consistent with those of \citet{gonze2002} who studied this model previously using the SSA, although their parameters values differ slightly from ours. Most notably, they consider $n=4$ rather than $n=2$ in the first term of Equation (\ref{eq:M_ODE}).\par 

The period obtained from the calibration simulation matches the value from the SSA, and the average concentrations and amplitudes are similar between the two, with the values from the calibration simulation being slightly higher. Because the calibration simulation uses periodic boundary conditions and the implicit enzymes are assumed to be well mixed, we do not expect to observe significant spatial effects. Therefore, the small differences between the concentrations likely arise from the use of the approximate reaction boundary.\par 

In contrast, the results of the spatial simulation, which implements reflective boundaries and explicitly models the nuclear membrane, differ significantly from those of the SSA and calibration simulations, highlighting the importance of spatial effects. Notably, the period is approximately one hour longer than in the calibration simulation, and the average amplitude of oscillations is higher for all reactants except $M$. In addition, the spatial simulation yields the highest average concentrations for all reactants.\par 

These discrepancies arise because reactions are less likely to occur near the boundaries of the domain than in the calibration simulation. Although the (implicit) enzymes are still well mixed, if a position near the boundary is sampled, then a reaction is less likely to occur compared to an enzyme position chosen far from the boundary. This is because the boundary reduces the volume that can be occupied by the nearest molecule of the relevant reactant. In \citeauthor{og_circadian_model}'s model, this effect impedes the degradation of $P_2$ (the final term in Equation (\ref{eq:P2_ODE}), which is the only mechanism for removing PER protein from the system. Consequently, the total PER protein concentration ($P_\text{total}$), is higher than in the calibration simulation. Similarly, the boundaries impede the production of $M$ in (the first term of Equation (\ref{eq:M_ODE})), which explains why the average value of $M$ only slightly exceeds the calibration simulation---while its average amplitude is actually the lowest of the three simulations---even though $P_\text{N}$ is highest in the spatial simulation. \par

The primary purpose of Section \ref{subsec:oscillations} is to demonstrate how our method can be applied to biochemical systems that appear commonly in literature. We emphasise that our method could be adapted to many other models, including to models of oscillations in NF$\kappa B$,  p$53$, or Wnt signalling cascades  \cite{doi:10.1073/pnas.87.4.1461, NFKB}. Although not demonstrated in this article, our reaction conditions can be used alongside Smoluchowski's original bimolecular condition, or its derivatives. Furthermore, all of the surface interactions studied by \citet{Andrews_2009} are independent of the reaction condition and are therefore compatible with our framework.\par 

The consequences of Assumption 3, while less obvious, are more severe. Typically, computing the inverse Laplace transform is not problematic. However, requiring that the resulting $f$ is non-negative limits the kinetics that can be simulated. Systems involving allosteric enzymes \cite{CARDENAS201510, cornish2014understanding} rarely obey Michaelis-Menten kinetics and are instead characterised by more general non-elementary reaction rates. Many examples exist, but in the present context the most relevant is the Hill-type rate that describes the production of $M$ in Equation (\ref{eq:M_ODE}). For $n=2$ the corresponding reaction probability given in Equation (\ref{eq:Hill_reaction_prob}) is valid, but for $n=4$ (originally considered by \citet{og_circadian_model}) we find
\begin{equation}
    F_{\text{H}4}(r_2) = 1-\text{cos}\left(\frac{4\pi r_2^3}{3\sqrt{K_I}}\right)\text{cosh}\left(\frac{4\pi r_2^3}{3\sqrt{K_I}}\right),
\end{equation}
which cannot be interpreted as a reaction probability since it is negative for some values of $r_2$.\par 

We could simply ignore the negative portion of the boundary---that is, set the reaction probability to $0$ when $F(r_2) < 0$---but in general this will cause the simulated rate to exceed the desired rate. It may be possible to correct for this by modifying the boundary in a manner analogous to that used to account for the additional flux over the boundary \cite{corrections_paper}. Alternatively, we could consider a reversible reaction where the forward and backward rates are governed by the positive and negative portions of the boundary, respectively. This scheme is not problematic in isolation, but in general the backwards reaction needs to be treated carefully to avoid its products causing artificial reactions due to them being initialised too close to other reactants \cite{Andrews_2004, klann2013reaction}. How best to circumvent such problematic boundaries remains an open question.\par

In general, the validity of non-elementary reaction rates in a statistic setting is system dependent. Non-elementary kinetics can be incorporated into the CME and simulated directly using the SSA, but only particular reaction rates, including those generated by Michaelis-Menten kinetics, correspond directly to propensity functions \cite{gillespie_MM_CME}. \citet{thomas2012} showed that the resulting SSA tends to overestimate the noise for Michaelis-Menten kinetics and underestimate it for Hill-type kinetics. In specific cases, this can even lead to the CME for the reduced system failing to predict noise-induced oscillations present in the full system.\par 

Despite these limitations, there are many systems for which the use of non-elementary kinetics is valid. For example, \citet{gonze2002} demonstrated that applying the SSA directly to \citeauthor{og_circadian_model}'s model yields dynamics consistent with stochastic simulations of the full system. Moreover, the full mechanistic details of many biological systems remain unknown and are instead described using reduced models that often incorporate non-elementary reaction rates. Therefore, tools like those developed here that provide insight into stochastic and spatial effects in such systems remain valuable.\par

Finally, we note that simulating non-elementary kinetics as demonstrated here cannot generally be replicated using the RDME, which is the primary alternative to particle-based models. This is because the RDME does not converge to the CME in the limit of fast diffusion when non-elementary propensities are used \cite{RDME_MM_nonconvergence}. The RDME is accurate in the limit of slow diffusion, since in this case it reduces to a collection of noncommunicating voxels each described by the CME. In contrast, in our approach, reactivity is already mediated by the CME and in the limit of fast diffusion, the reactants become well-mixed globally, which ensures Assumption 1 holds. Therefore, as long as an appropriate reaction boundary can be determined and assuming that the errors that arise from adopting this boundary are negligible, we expect to recover the CME. Moreover, in the limit of slow diffusion, our method should also be accurate assuming that the reactants are well-mixed on a spatial scale sufficient to satisfy Assumption 1. Notice that this is not dissimilar to the RDME which requires reactants to be well-mixed within individual voxels.\par 

\section{Conclusion}

In this article, we have developed and validated a particle-based simulation framework capable of directly reproducing non-elementary bimolecular kinetics without explicitly simulating the underlying sequence of elementary reactions. Through the conceptual use of a phantom reactant we demonstrate how to adapt the reaction boundaries proposed by \citet{our_first_paper} for trimolecular systems to bimolecular reactions. Thereby enabling the simulation of typical non-elementary kinetics including those of Michaelis-Menten and Hill.\par

The accuracy and utility of our method was shown using two biologically relevant examples. In Section \ref{subsec:MM_results}, we applied our method to a prototypical Michaelis–Menten system and demonstrated excellent agreement between the simulated and theoretical rates. In Section \ref{subsec:oscillations}, we simulated a minimal model of circadian oscillations, which includes multiple non-elementary reactions. By combining our approach with standard particle-based techniques, we are able to simulate the full system without reverting to the underlying elementary reactions. Our results are similar to those obtained using the CME when periodic boundary conditions are applied to the domain, but differ noticeably when reflecting boundary conditions are implemented, demonstrating the importance of spatial effects in cellular systems.\par

Our method broadens the class of kinetics that can be directly simulated in particle-based models and facilitates study of spatial and stochastic dynamics in models that incorporate non-elementary reactions in a manner that cannot be replicated by the RDME. It is straightforward to implement programmatically, can be easily incorporated alongside existing methods, and enhances the efficiency of particle-based models for enzymatic systems. 

\newpage

\printbibliography

\end{document}